\documentclass{pasj01}
\draft 
\Received{$\langle$2022 May 26$\rangle$}
\Accepted{$\langle$2022 Aug 09$\rangle$}
\Published{$\langle$publication date$\rangle$}

\usepackage{lscape}

\newcommand{\tabref}[1]{table~\ref{#1}}
\newcommand{\equref}[1]{eq~(\ref{#1})}
\newcommand{\figref}[1]{figure~\ref{#1}}
\newcommand{\Figref}[1]{Figure~\ref{#1}}

\usepackage[hang,small,bf]{caption}
\usepackage[subrefformat=parens]{subcaption}
\begin{document}

\title{ Starspots, chromospheric emission lines, and flares of zero-age main-sequence stars }
\author{Mai Yamashita${}^1$, Yoichi Itoh${}^1$, Yumiko Oasa${}^2$}
\altaffiltext{}{${}^1$Nishi-Harima Astronomical Observatory, Center for Astronomy, University of Hyogo, 407-2 Nishigaichi, Sayo, Sayo, Hyogo 679-5313 }
\altaffiltext{}{${}^2$Faculty of Education, Saitama University, 255 Shimo-Okubo, Sakura, Saitama, Saitama, Japan}
\email{yamashita@nhao.jp}


\maketitle

\begin{abstract}
Zero-age main-sequence (ZAMS) stars are considered to have enormous starspots and show strong chromospheric emission lines because of their strong surface magnetic field. We discuss the dynamo activities of ZAMS stars with respect to their periodic light variation caused by a starspot and with respect to the strength of the chromospheric emission lines. The light curves of $33$ ZAMS stars in IC 2391 and IC 2602 were obtained from {\it TESS} photometric data. The light curves can be grouped into the following four categories: single frequency, possible shape changer, beater, and complex variability. The amplitudes of the light curves are $0.001-0.145\,\mathrm{mag}$, similar to those of ZAMS stars in Pleiades. The starspot coverages are $0.1-21\%$. 
We found that the light variations and Ca I\hspace{-.1em}I emission line strength of ZAMS stars in IC 2391, IC 2602, and the Pleiades cluster are as large as those of the most active superflare stars and two orders larger than those of the Sun, and are located on the extensions of the superflare stars. These results suggest that superflare stars link the properties of the Sun to those of the ZAMS stars of ages between $30$ and $120\,\mathrm{Myr}$. 
ZAMS stars with a single frequency or possible shape change in the light curve tend to have both large light variation, indicating large spot coverage, and saturated Ca I\hspace{-.1em}I emission line strength. ZAMS stars with beat or complex variability have small spot coverage and a faint Ca I\hspace{-.1em}I emission line.
We also detected $21$ flares in the {\it TESS} light curves of $12$ ZAMS stars in IC 2391 and IC 2602, where most of these stars have saturated chromospheric Ca\, \emissiontype{II} emission lines. The energies of the flares are estimated to be $\sim 10^{33}-10^{35}\,\mathrm{erg}$, which is comparable with the energy of a superflare. 
\end{abstract}


\section{Introduction} 
\label{intro}

The chromosphere is the active atmosphere in which atoms emit some permitted lines, such as H${\rm \alpha}$ and Ca\,\emissiontype{II}. It is claimed that chromospheric activity is driven by the magnetic fields induced by the dynamo process. In the dynamo process, the Coriolis force (= rotational moment $\times$ convection velocity) balances the Lorentz force (= current $\times$ magnetic strength / density of plasma) \citep{ba96}. Stellar rotation and convection are considered to be the main processes that drive the evolution of magnetic activities. \citet{noyes} used the Rossby number, $N_{\rm R}$, as an indicator of magnetic activity. It is defined as $P/\tau_{\rm c}$, where $P$ is the stellar rotational period and $\tau_{\rm c}$ is the convective turnover time. $N_{\rm R}$ can be approximated as the inverse square of the dynamo number, $N_D$, the wave solution of the dynamo equation. Magnetic fields develop when $|N_D| > 1$. 

Chromospheric emission lines are observational evidence of a strong magnetic field. The relationship between chromospheric line strength and the Rossby number has been examined for zero-age main-sequence (ZAMS) stars. \citet{m09} detected Ca\,\emissiontype{II} infrared triplet (IRT; $\lambda 8498, 8542, 8662 \, \mathrm{\AA}$) emission lines from low-mass stars in the young open clusters IC 2391 and IC 2602. The cluster members are considered to be on ZAMS or in the last evolution phase prior to ZAMS. They calculated $R^{\prime}_{\rm IRT}$ from the equivalent widths (EQWs). $R^{\prime}_{\rm IRT}$ describes the ratio of the surface flux of the Ca\,\emissiontype{II} IRT emission lines to the stellar bolometric luminosity. They found that $R^{\prime}_{\rm IRT}$ decreases with increasing $N_{\rm R}$ for stars with $N_{\rm R} \geq 10^{-1.1}$. This region is called the unsaturated regime. In contrast, $R^{\prime}_{\rm IRT}$ is constant at levels of approximately $10^{-4.2}$ for stars with $N_{\rm R} \leq 10^{-1.1}$. This region is called the saturated regime. \citet{m09} suggested that the chromosphere is completely filled by the emitting regions for the stars in the saturated regime. Recently, \citet{f21} found that some of the FGKM-type stars in the young open cluster NGC 3532 ($300\,\mathrm{Myr}$) also show saturation at $R^{\prime}_{\rm IRT} \sim 10^{-3.7}$. \citet{y20} investigated the relationship between $N_{\rm R}$ and the Ca\,\emissiontype{II} IRT emission lines of $60$ pre-main sequence (PMS) stars. Only three PMS stars showed broad and strong emissions, indicative of large mass accretion. Most of the PMS stars presented narrow and weak emissions, suggesting that their emission lines are formed in the chromosphere. All their Ca\,\emissiontype{II} IRT emission lines have $R^{\prime}_{\rm IRT} \sim 10^{-4.2}$, which is as large as the maximum $R^{\prime}_{\rm IRT}$ of ZAMS stars. The PMS stars show $N_{\rm R} < 10^{-0.8}$ and constant $R^{\prime}_{\rm IRT}$ against $N_{\rm R}$, i.e., their Ca\,\emissiontype{II} IRT emission lines are saturated. \citet{y22} (hereafter Paper I) investigated the infrared Mg\, \emissiontype{I} emission lines at $8807 \, \mathrm{\AA}$ of 47 ZAMS stars in IC 2391 and IC 2602 using the archive data of the Anglo-Australian Telescope at the University College London Echelle Spectrograph. They found that ZAMS stars with smaller Rossby numbers show stronger Mg\, \emissiontype{I} emission lines, even for stars located in the Ca\,\emissiontype{II} saturated region. 

Regarding the solar disk, it is known that spots are surrounded by emission regions, such as faculae in the photosphere and plages in the chromosphere. Other observational evidence for the strong magnetic field of a star includes the periodic light variation caused by a starspot. The amplitude of the solar brightness variation caused by sunspots on the rotating solar surface is $0.01-0.1\%$ \citep{l03}. 
In recent years, \textit{Kepler} and Transiting Exoplanet Survey Satellite (\textit{TESS}, Ricker et al. 2015) satellites have been utilized to investigate stellar magnetic activities. The typical photometric precision of \textit{Kepler} is $0.01\%$ for a star of $12 \,\mathrm{mag}$ \citep{k10}. The time resolution of \textit{Kepler} is approximately $30 \,\mathrm{minutes}$ and $1\,\mathrm{minutes}$. \citet{r15a} presented \textit{K2} light curves for F-, G-, K-, and M-type ZAMS stars in the Pleiades cluster. The rotational periods ranged from $0.082\, \mathrm{d}$ to $22.14 \, \mathrm{d}$. The amplitude of the brightness ranged from $0.001\, \mathrm{mag}$ to $0.556 \, \mathrm{mag}$. \citet{r15a} and \citet{st16} investigated the relationship between the rotational periods and shapes of the light curves for ZAMS stars in the Pleiades cluster and found that approximately half of the F-, G-, and K-type ZAMS stars rotating rapidly have sinusoidal light curves. 

Flares are also observational evidence for a strong magnetic field. \citet{m13} detected $365$ flares with \textit{Kepler} data regarding $148$ solar-type stars. The releasing energy of the flare is $100$ times larger than those of solar flares. Such objects are called superflare stars. \citet{n15} found that the Sun and superflare stars show a positive correlation between the amplitude of the light curve and the $r_0$(8542) index, the residual core flux of the Ca\,\emissiontype{II} IRT normalized by the continuum level at the line core. This correlation means that superflare stars have large starspots and high magnetic activity compared to the Sun. \citet{i21} detected $3844$ flares from \textit{K2} data from $2111$ objects associated with open clusters. The cluster ages range from $0.1\,\mathrm{Gyr}$ to $3.6\,\mathrm{Gyr}$. They found that flaring rates decline with age and decline faster for stars with higher mass. 

We discuss the relationship between the periodic light variation caused by a starspot, the strength of the chromospheric emission lines, and flare events. We measured the light variations of $33$ ZAMS stars in IC 2391 and IC 2602 with {\it TESS} photometric data. In the next section, we describe the photometric data reduction procedure. In Section \ref{result}, we present the results, and in Section \ref{discussion}, we discuss the dynamo activities in terms of the light variation, the Rossby number, the chromospheric emission strength, and the flares.

\section{Data Sets and Data Reduction}
\label{observ}

\subsection{Stellar parameters}
Our targets are F-, K-, and G-type ZAMS stars in IC 2391 ($50 \pm 5 \, \mathrm{Myr}$; \cite{ba04}) and IC 2602 ($30 \pm 5 \, \mathrm{Myr}$; \cite{st97}). The metallicity of both clusters has been determined to be similar to that of the Sun, which is ${\rm [Fe/H]} = -0.01\pm 0.02$ for IC 2391 \citep{d09} and ${\rm [Fe/H]} = 0.00 \pm 0.01$ for IC 2602 \citep{ra01}. A total of $44$ ZAMS stars have been confirmed as single stars in the IC 2391 and IC 2602 members based on the strength of the lithium $6708\, \mathrm{\AA}$ absorption line \citep{m09}. In Paper I, we examined the proper motion and radial velocity of those stars, and five ZAMS stars were removed from our target list. The 39 targets investigated in this study are presented in \tabref{tab:zams1}. 

\subsection{Data Reduction}
The members of IC 2391 and IC 2602 were observed in the long-cadence ($1800$-s exposure) mode of {\it TESS} Sectors 8 and 10, which extended from the 2nd to the 27th of February 2019, and from the 26th of March 2019 to the 21st of April 2019, respectively. These data were retrieved from the Multimission Archive at the Space Telescope Science Institute. We conducted principal component analysis using ELEANOR, an open-source tool for extracting light curves from {\it TESS} full-frame images \citep{f19}. ELEANOR enables us to model a 2D background, and remove the extended point-spread function of bright stars across the {\it TESS} images. However {\it TESS} pixels are substantially larger than that of \textit{Kepler}. We excluded 6 ZAMS stars from the targets because another star whose magnitude is brighter than $< 3 + V\,\mathrm{mag}$ of the ZAMS star is located in the aperture of the ZAMS star. 

First, we search for periodic signals and obtain the amplitudes of the light curves. We calculated the standard deviation of the flux, $\sigma_1$, and removed the flux data points greater than $3\sigma_1$ above the average. As described in Paper I, we searched for periodic signals by conducting Lomb--Scargle periodogram analysis \citep{s82}. We note that VXR PSPC 44 showed beating signatures in its light curve; we adopted the period  for the second maximum of the power as the period of the object. We calculated the amplitudes of the light curves by taking the 90th percentile flux and subtracting it from the 10th percentile flux. $P$ was judged to be accurate if ${\rm amplitude/mean \, flux \, error} \geq 10$. The periods derived by us show good correlation with the periods measured by \citet{ps96} and \citet{bs99}, with a correlation coefficient of $0.997$. 

\begin{figure}[htbp] \hspace{-10mm}
 \begin{minipage}{0.74\hsize}
  \begin{center}
   \includegraphics[height=51mm]{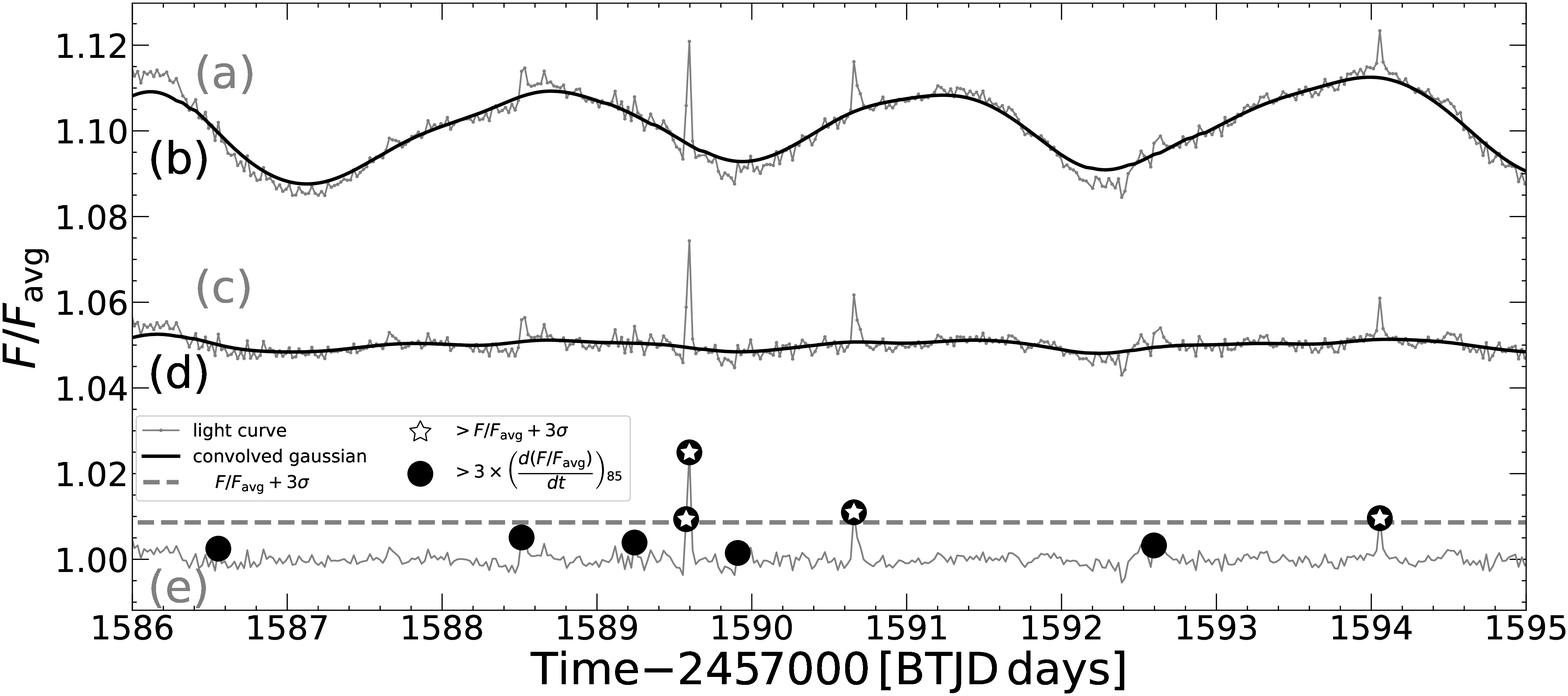}
  \end{center} 
 \end{minipage}\hspace{-11mm}
 \begin{minipage}{0.24\hsize}
  \begin{center}
   \includegraphics[height=51mm]{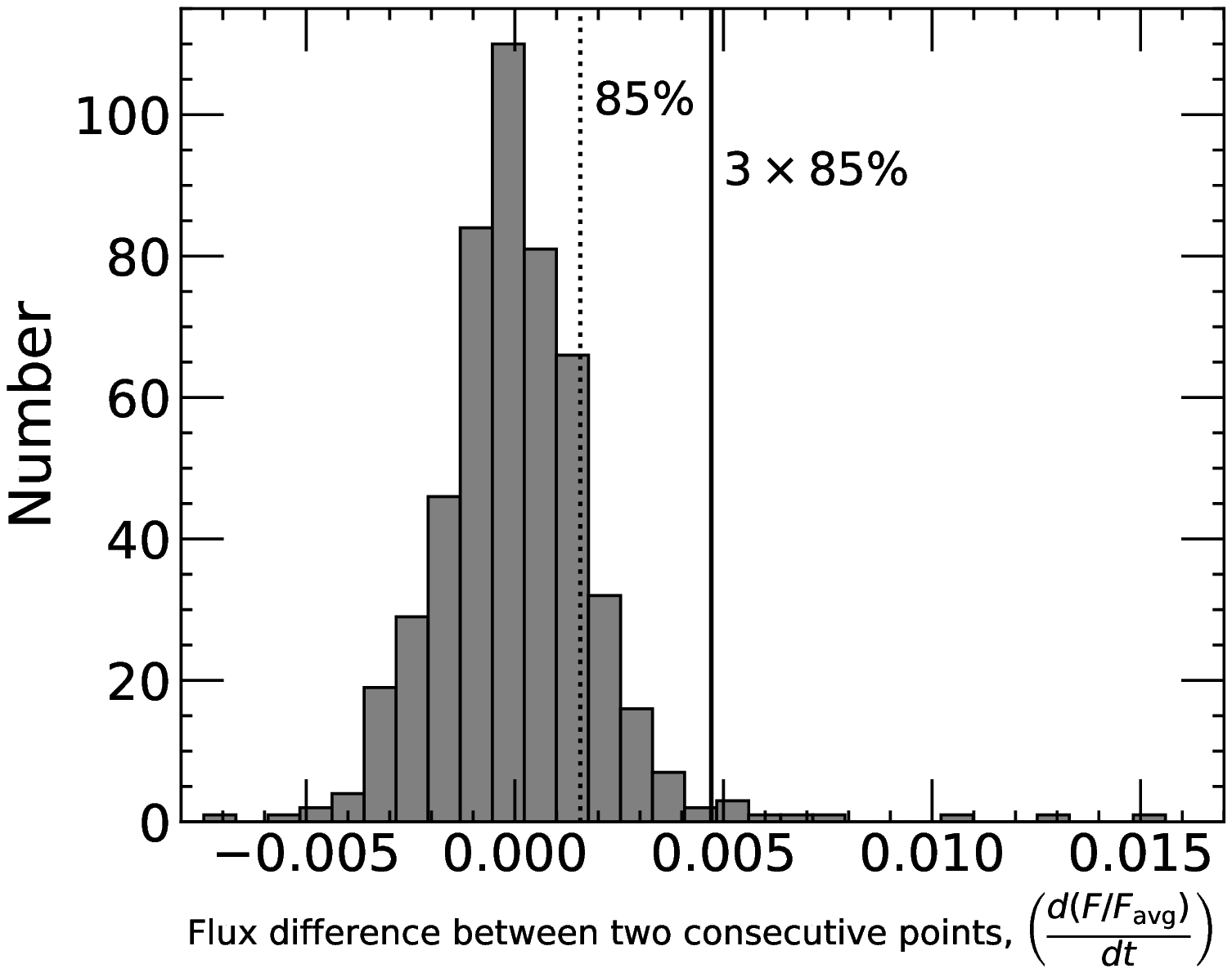}
  \end{center}
 \end{minipage}
 \caption{Light curve of the ZAMS star, [RSP 95] 14. Horizontal and vertical axes correspond to days and normalized flux. Top) The original light curve is shown with the gray solid line (a). This is divided by the median of the counts and normalized to $1$. The thick black line is the Gaussian convolved light curve (b). The standard deviation of the Gaussian function is set to $300\,\mathrm{mins}$. These lines are shifted by $+0.10$ for display purposes. Middle) The ratio between the original light curve and the Gaussian convolved light curve shown in the top panel is presented with the gray solid line (c). The thick black line is the Gaussian convolved light curve of the gray solid line (d). The standard deviation of the Gaussian function is set to $300\,\mathrm{mins}$. These lines are shifted by $+0.05$ for display purposes. Bottom) The ratio between the light curve shown in the middle and the second Gaussian convolved light curve is shown with the gray solid line (e). The dashed line denotes $1+3\sigma$ of the gray line. The open symbols represent data points greater than the dashed line. The filled circles represent data points whose brightness change is greater than the threshold of the change of the brightness. The three sudden brightenings (four points) are detected as flares. The histogram in the right panel shows the number distribution of the flux differences between all pairs of two consecutive data points. The dotted line shows the top 15\% of the distribution. The solid line shows the threshold of the change of the brightness, three times in the top 15\% of the distribution. } \label{fig:flare_detection}
\end{figure}

Next, we detected sudden brightenings as flares. Starting from the original light curves (including $3\,\sigma_1$ outliers), first we removed the rotational light variations as described below (\figref{fig:flare_detection}). \textit{TESS} data was not obtained for several days in the middle of the observation period. We separated the light curve at the center time. Each light curve was divided by the median of the counts and normalized to $1$ (\figref{fig:flare_detection} (a)). Each light curve was convolved with a Gaussian function (\figref{fig:flare_detection} (b)) and divided by the convolved light curve. The derived light curve (\figref{fig:flare_detection} (c)) was again convolved with a Gaussian function (\figref{fig:flare_detection} (d)). We set the standard deviation of the Gaussian function to between $15\,\mathrm{mins}$ and $2.5\,\mathrm{days}$ for each light curve. With these processes, no rotational light variation remained in the light curve, and most of the data points were set around unity (\figref{fig:flare_detection} (e)). We then measured the standard deviation of the counts in the light curve, $\sigma_2$, for each ZAMS star. 

We defined two thresholds for flare detection. The first threshold is the height of sudden brightening. After removing the rotational light variations, we searched for sudden brightenings whose relative flux exceeded $1+3\sigma_2$ for at least one data point. The second threshold is the change of the brightness. We calculated the brightness changes between all pairs of consecutive data points. We searched for any brightness change that occurred more than three times in the top 15\% of the distribution (the solid line in the right panel of \figref{fig:flare_detection}). We identified a sudden brightening that exceeded both thresholds as a flare. The time at which the flux first exceeded the first threshold, $1+3\sigma_2$ of the relative flux, was determined to be the flare start time. The flare end time was defined as the time at which the flux became smaller than $1+3\sigma_2$ of the relative flux.

\begin{small}
\renewcommand{\tabcolsep}{4pt}  
\begin{longtable}{lccrrcrclcc} 
\caption{Physical parameters of ZAMS stars in IC 2391 and IC 2602.}
\label{tab:zams1}    
\hline\noalign{\vskip3pt} 
Object name & RA&Dec& $V$     & $K$   & Period  & $\log\, N_{\rm R}$    & $\log\,R^{\prime}_{\lambda 8542}$ & Amp & Spot coverage  & LC type \\
 & $\mathrm{[deg]}$ & $\mathrm{[deg]}$ & $\mathrm{[mag]}$   & $\mathrm{[mag]}$    & $\mathrm{[days]}$ &        &              & $\mathrm{[mag]}$           &   &         \\
(1)  &  (2) &  (3) &  (4) &  (5) &  (6) & (7) &  (8) &  (9) & (10)  & (11)\\    [2pt] 
\hline\noalign{\vskip3pt} 
\endhead
\endfoot 
\multicolumn{2}{@{}l@{}}{\hbox to0pt{\parbox{165mm}{\footnotesize 
{References for parameters. (4) $V$-mag: \citet{m09}. }{(5) $K_{\rm s}$-mag: 2MASS survey \citep{c03}. }{(6) Period: \citet{y22}, ${}^a$\citet{ps96} and ${}^b$\citet{bs99}. }{(7)(8)(9) The Rossby number $N_{\rm R}$, the ratio of the surface flux of the Ca\,\emissiontype{II} IRT emission line at $\lambda 8542$ to the stellar bolometric luminosity $R^{\prime}_{\lambda 8542}$, and the amplitudes of the light curve: \citet{y22}. }
}\hss}} 
\endlastfoot
IC 2391        &    &       &    &        &                &        &              &               &       &         \\ \hline
Cl* IC2391 L32 & 130.57590 & -53.90227 & 9.38  & 8.325  &- & -0.31 & -4.84 & 0.002 & 0.002 & III   \\
VXR PSPC 07    & 129.74920 & -53.02395 & 9.63  & 8.272  & - & 0.04  & -6.04 & 0.002 & 0.002 & III   \\
VXR PSPC 12    & 129.97104 & -52.96579 & 11.86 & 9.793  & 3.69 & -0.90 & -4.34 & 0.030 & 0.039 & I    \\
VXR PSPC 22A   & 130.20459 & -53.62927 & 11.08 & 9.272  & 2.31 & -0.88 & -4.34 & 0.025 & 0.030 & I    \\
VXR PSPC 44    & 130.55122 & -53.10105 & 9.69  & 8.364  & 0.57 & -0.36 & -4.62 & 0.009 & 0.010 & II   \\
VXR PSPC 45A   & 130.56148 & -52.93398 & 10.70 & 8.648  & 0.22 & -2.12 & -4.08 & 0.020 & 0.026 & I    \\
VXR PSPC 52    & 130.69420 & -53.01704 & 10.34 & 8.991  & 2.15 & -0.79 & -4.70 & 0.013 & 0.016 & I'   \\
VXR PSPC 66    & 130.96793 & -53.23327 & 9.73  & 8.594  & 0.92 & -0.42 & -4.98 & 0.004 & 0.005 & I'   \\
VXR PSPC 67A   & 130.98671 & -52.68489 & 11.71 & 9.424  & 3.41 & -0.93 & -4.37 & 0.014 & 0.018 & I    \\
VXR PSPC 69A   & 130.99590 & -53.56213 & 11.67 & 9.675  & 2.22 & -1.12 & -4.24 & 0.098 & 0.131 & I    \\
VXR PSPC 77A   & 131.41306 & -52.43321 & 9.91  & 8.635  & 0.65 & -1.11 & -4.56 & 0.019 & 0.022 & I    \\ \hline

IC 2602        &    &       &    &        &                &        &              &               &       &         \\ \hline
Cl* IC2602 W79 & 160.52952 & -64.76884 & 11.57 & -  & 6.55 & -0.73 & -4.57 & 0.011 & 0.014 & III   \\
RSP95 1        & 157.13011 & -63.73767 & 11.57 & 9.537  & 3.85 & -0.96 & -4.33 & 0.026 & 0.033 & I'   \\
RSP95 7        & 157.93717 & -63.48943 & 9.21  & 8.212  & - & -0.11 & -5.00 & 0.001 & 0.001 & III   \\
RSP95 10       & 158.11917 & -65.10722 & 12.77 & -  & 3.16 & -1.18 & -4.21 & 0.068 & 0.099 & I'   \\
RSP95 14       & 158.41530 & -64.78116 & 11.57 & 9.572  & 2.73 & -1.11 & -4.25 & 0.020 & 0.025 & I'   \\
RSP95 29       & 159.15815 & -64.79828 & 12.73 & 9.976  & 2.22 & -1.34 & -4.20 & 0.054 & 0.074 & I    \\
RSP95 35       & 159.57360 & -64.13512 & 10.59 & 9.078  & 2.46 & -0.71 & -4.32 & 0.018 & 0.021 & I'   \\
RSP95 43       & 159.98321 & -63.99169 & 12.14 & 9.622  & 0.78 & -1.66 & -3.92 & 0.103 & 0.141 & I    \\
RSP95 52       & 160.21399 & -64.71329 & 12.19 & 9.555  & 0.39 & -1.95 & -4.12 & 0.145 & 0.206 & I    \\
RSP95 58       & 160.67298 & -64.35120 & 10.52 & 8.840  & 0.57 & -1.57 & -4.18 & 0.038 & 0.046 & I    \\
RSP95 59       & 160.73377 & -63.93087 & 11.86 & 9.537  & 1.31 & -1.43 & -4.16 & 0.066 & 0.089 & I    \\
RSP95 66       & 161.02838 & -63.99307 & 11.07 & 9.311  & 3.28 & -0.80 & -4.46 & 0.028 & 0.035 & I'   \\
RSP95 68       & 161.05806 & -64.77426 & 11.32 & 8.750  & 0.99 & -1.55 & -4.05 & 0.038 & 0.050 & I'   \\
RSP95 70       & 161.09396 & -64.25837 & 10.92 & 9.322  & 4.25 & -0.47 & -4.59 & 0.018 & 0.021 & I    \\
RSP95 72       & 161.24842 & -65.03863 & 10.89 & 9.157  & 1.05 & -1.30 & -4.15 & 0.048 & 0.060 & I    \\
RSP95 79       & 161.36767 & -64.22918 & 9.08  & 7.989  & 0.75 & -0.24 & -4.73 & 0.003 & 0.003 & I'   \\
RSP95 80       & 161.37445 & -64.42227 & 10.66 & 8.278  & 7.25 & -0.69 & -4.54 & 0.013 & 0.018 & III   \\
RSP95 83       & 161.56180 & -64.04946 & 10.7  & 9.102  & 1.74 & -1.08 & -4.37 & 0.008 & 0.010 & I'   \\
RSP95 85       & 161.61995 & -64.13266 & 9.87  & 8.680  & 1.33 & -0.80 & -4.91 & 0.003 & 0.004 & I'   \\
RSP95 89       & 161.71584 & -63.57105 & 12.97 & 9.942  & 4.73 & -1.01 & -4.26 & 0.081 & 0.119 & I    \\
RSP95 92       & 162.07663 & -64.16482 & 10.26 & 8.505  & 1.93 & -0.82 & -4.42 & 0.024 & 0.028 & I   \\
RSP95 95A      & 162.45167 & -64.77457 & 11.73 & 9.480  & 1.22 & -1.46 & -4.20 & 0.007 & 0.009 & I   \\ \hline \\
\end{longtable} 
\renewcommand{\tabcolsep}{6pt}
\end{small}

\clearpage
\section{Results}
\label{result}

The rotational periods of the ZAMS stars in IC 2391 and IC 2602 range from $0.22 - 7.25\,\mathrm{days}$. The amplitudes of the light curves range from $0.001-0.145\,\mathrm{mag}$. The measured periods and amplitudes are similar to those of $88$ ZAMS stars in Pleiades \citep{r15a}. Previous ground-based observation \citep{m11} also detected a variation of $0.2\,\mathrm{mag}$ in the brightness of 8 ZAMS stars in IC 2391 (namely, VXR PSPC 12, VXR PSPC 14, VXR PSPC 22A, VXR PSPC 35A, VXR PSPC 69, VXR PSPC 70, VXR PSPC 72, VXR PSPC 76A). 

\begin{figure}[htb]
	\centering
	\includegraphics[width=16cm]{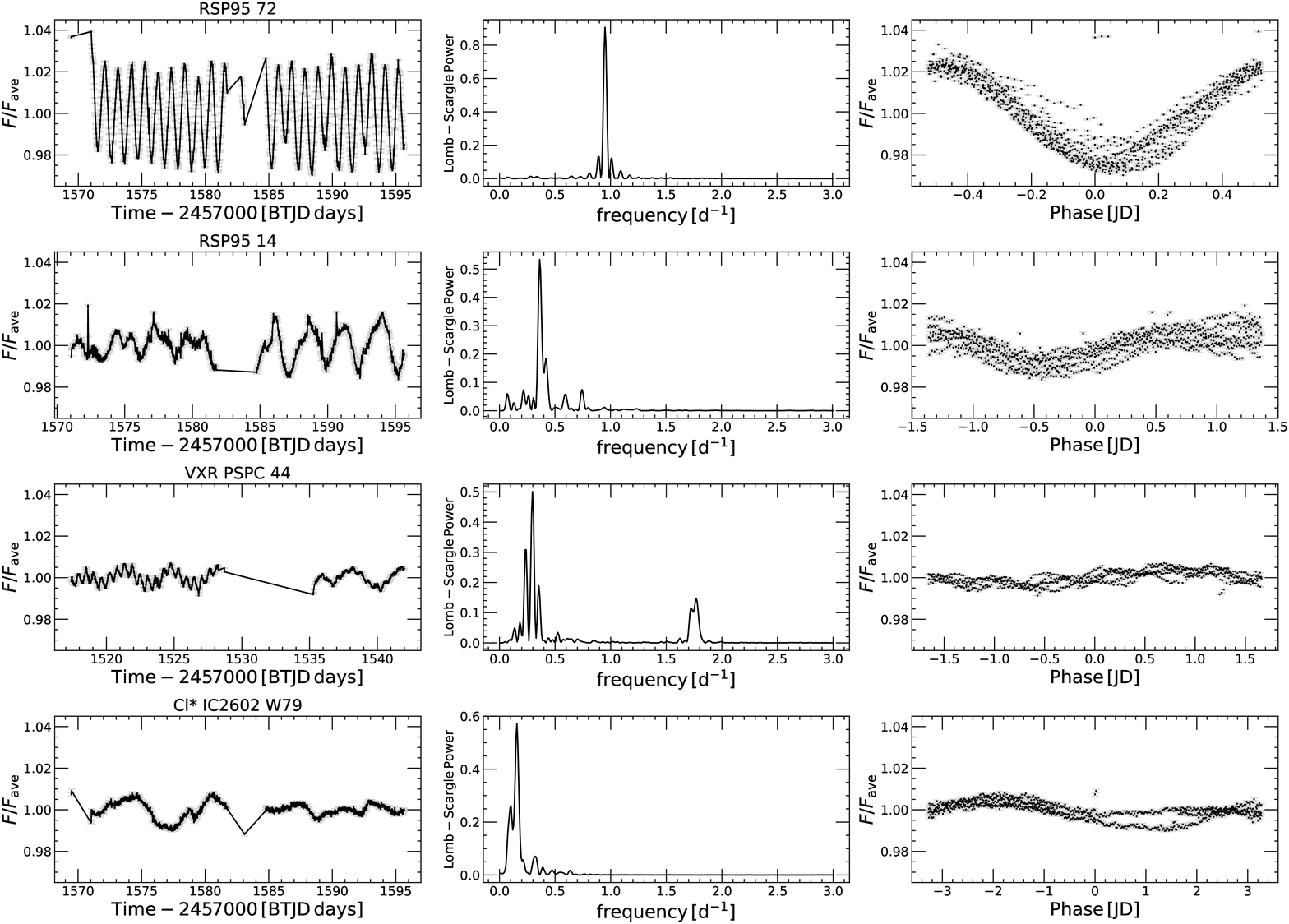} \caption{Four examples of the light curves of ZAMS stars. The four features of the light curve, I) single frequency, I') possible shape changer, II) beater, and III) complex variability, are presented in order from the top to the bottom. Left column: full light curve; middle column: Lomb-Scargle periodogram; right column; phased light curve. Data points greater than $3\sigma_1$ above the average flux were removed from the phased light curve. } \label{fig:t1} 
\end{figure}

\Figref{fig:t1} shows four examples of light curves. We categorized the light curves into the following groups: 
	\begin{enumerate}
	\item[I)] Single frequency: the star has a single period. A single spot or spot group is interpreted as rotating into and out of view.
	\item[I')] Possible shape changer: subcategory of type I) single frequency. The star usually has a single period. Shape of the light curve changes over the campaign, but not enough such that a separate period can be derived. 
	\item[II)] Beater: the light curve appears to have a beating signature. The star has multi periods.  Spot or spot group evolution and/or strong latitudinal differential rotation are suggested. 
	\item[III)] Complex variability; the light curve has periodicity but seems complex. It is considered that the star has multiple sunspot groups. 
	\end{enumerate}
Single frequency and beater are all classes found in \citet{r15b}. For the 33 ZAMS stars in IC 2391 and IC 2602, we categorized $16$ ZAMS stars as type I) single frequency, $11$ ZAMS stars as type I') possible shape changer, $1$ ZAMS star as type II) beater, and $5$ ZAMS stars as type III) complex variability (\tabref{tab:lc}). 

The apparent area of the spot, $A_{\rm spot}$, was estimated from the amplitude of light variation ($\Delta F/F$) using the following equation (Notsu et al. 2013; Shibata et al. 2013): 
\begin{equation}
A_{\rm spot} = \frac{\Delta F}{F} A_{\rm \frac{1}{2}  star} \left[1- \left(\frac{ T_{\rm spot} }{ T_{\rm eff} } \right)^4 \right]^{-1}, 
\label{eq:spot}
\end{equation}
where $A_{\rm \frac{1}{2} star}$ is the effective area of the stellar hemisphere. $T_{\rm spot}$ and $T_{\rm eff}$ are the temperature of the starspot and photosphere of the star. We applied eq. (4) of \citet{n19}. It is an empirical equation on the temperature difference between photosphere and spot, $T_{\rm eff} - T_{\rm spot}$, deduced from \citet{b05}. 

The starspot coverages, $ A_{\rm spot} / A_{\rm \frac{1}{2} star} $, are estimated to be $0.1-21\%$ for the ZAMS stars in IC 2391 and IC 2602. The mean coverage of each group is listed in \tabref{tab:lc}. ZAMS stars with single frequency have the largest mean spot area, and possible shape changer ZAMS stars have the second largest values. The mean spot area of the beater and complex variability ZAMS stars are as same as the Solar value; $\lesssim 1\%$.

\begin{table}[htb]
  \centering
  \caption{Light curve groups and mean spot coverage of the $33$ ZAMS stars}
  \begin{tabular}{rllllll} \hline
\multicolumn{2}{l}{Features of the light curve} & \multicolumn{2}{c}{Number} & {} & \multicolumn{2}{c}{Mean spot coverage [\%]} \\ \cline{3-4} \cline{6-7} 
{} & {} & IC 2391 & IC 2602 &{} & IC 2391 & IC 2602  \\ \hline
I )  & Single frequency & 6 & 10 &{} & $4.43\pm4.33$ & $7.93\pm6.15$\\
I' ) & Possible shape changer & 2 & 9 &{} & $1.01\pm0.79$ & $3.12\pm2.97$\\
II ) &  Beater & 1 & 0 &{} & $1.00$ & - \\
III ) & Complex variability & 2 & 3 &{} & $0.19\pm0.03$ & $1.09\pm0.86$\\
 \hline
  \end{tabular}
\label{tab:lc}
\end{table}

The middle column of \figref{fig:t1} shows examples of Lomb-Scargle periodograms. The FAP for the strongest peaks was calculated to be $3\times 10^{-290} - 0.007$. \textit{TESS} data was not obtained for several days in the middle of the observation period. This leads to aliasing on the side of the strongest peak in the periodogram. When we split the light curve at the center time and applied Lomb--Scargle periodogram analysis, the aliasing disappeared. 

\begin{table}[htb]
\caption{ List of detected flares} \label{tab:flare}
\centering
\begin{tabular}{lclllrl}
\hline
Object name  & $N_{\rm flare}$ & Flare rate  & start & end & Duration time  & $E_{\rm flare}$   \\ 
  &  & $\,\mathrm{[d^{-1}]}$ & \multicolumn{2}{l}{$-2457000\,\mathrm{[BTJD days]}$} &  $\,\mathrm{[s]}$ & $\,\mathrm{[erg]}$   \\ 
(1)  &  (2) &  (3) &  (4) &  (5) &  (6) & (7) \\ \hline
IC 2309      &        &           &              &            &                           &                    \\ \hline
VXR PSPC 7  & 1 & 0.061 & 1527.732 & 1527.911 & 15500 & 2.79$\times 10^{34}$   \\
VXR PSPC 22A & 2 & 0.119 & 1521.507 & 1521.691 & 15900 & 1.10$\times 10^{34}$   \\
             &   &       & 1538.712 & 1538.893 & 15600 & 3.57$\times 10^{34}$   \\
VXR PSPC 67A & 1 & 0.066 & 1522.810 & 1523.031 & 19100 & 4.99$\times 10^{33}$   \\ \hline
IC 2602      &        &           &              &            &                           &                    \\ \hline
RSP95 14     & 4 & 0.211 & 1572.245 & 1572.394 & 12900 & 9.19$\times 10^{33}$   \\
             &   &       & 1589.550 & 1589.651 & 8700  & 9.09$\times 10^{33}$   \\
             &   &       & 1590.634 & 1590.788 & 13300 & 4.49$\times 10^{33}$   \\
             &   &       & 1594.010 & 1594.113 & 8900  & 2.39$\times 10^{33}$   \\
RSP95 43     & 1 & 0.049 & 1576.154 & 1576.233 & 6800  & 1.09$\times 10^{33}$   \\
RSP95 59     & 1 & 0.049 & 1595.024 & 1595.061 & 3200  & 3.59$\times 10^{33}$   \\
RSP95 66     & 1 & 0.049 & 1571.844 & 1571.953 & 9400  & 6.49$\times 10^{33}$   \\
RSP95 68     & 5 & 0.241 & 1578.423 & 1578.485 & 5400  & 6.00$\times 10^{33}$   \\
             &   &       & 1579.994 & 1580.064 & 6000  & 9.59$\times 10^{33}$   \\
             &   &       & 1580.417 & 1580.671 & 21900 & 3.79$\times 10^{33}$   \\
             &   &       & 1585.552 & 1585.652 & 8600  & 1.55$\times 10^{34}$   \\
             &   &       & 1586.795 & 1586.872 & 6700  & 1.69$\times 10^{33}$   \\
RSP95 72     & 1 & 0.050 & 1575.538 & 1575.639 & 8700  & 1.66$\times 10^{34}$ \\
RSP95 83     & 1 & 0.048 & 1586.471 & 1586.517 & 4000  & 3.19$\times 10^{33}$   \\
RSP95 89     & 1 & 0.050 & 1578.693 & 1579.124 & 37200 & 1.32$\times 10^{34}$ \\
RSP95 95A    & 2 & 0.096 & 1571.429 & 1571.710 & 24300 & 2.10$\times 10^{34}$   \\
             &   &       & 1588.191 & 1588.419 & 19700 & 2.13$\times 10^{34}$   \\ \hline
\end{tabular}
\end{table}

We also detected $21$ flares in the light curves of $12$ ZAMS stars in IC 2391 and IC 2602. The properties of each flare are listed in \tabref{tab:flare}. Their peak amplitude and duration time range from $0.3\%$ to $5.2\%$ and from $3200\,\mathrm{s}$ to $37200\,\mathrm{s}$, respectively. For the case of the first flare shown in \figref{fig:flare_detection}, the relative flux and the duration are $2.0\%$ and $2.4\,\mathrm{h}$, respectively. As we will discuss in section \ref{flare}, the total bolometric energy of the flare is estimated to be $\sim 10^{33}-10^{35}\,\mathrm{erg}$. It is comparable of that of superflares detected in \citet{m13}, in which the energy of the superflare is calculated to be $100$ times larger than those of solar flares, i.e. $10^{33}$ to $10^{36}\,\mathrm{erg}$. The occurrence frequency of flares can be estimated from the number of observed flares, the number of observed stars, and the length of the observation period. In this study, $21$ flares were detected from the $33$ ZAMS stars in the observational period, so that the occurrence frequency of flares is calculated to be $10.9$ flares per year per star. \citet{m13} detected $14$ superflares from data collected from approximately $14,000$ slowly rotating G-type main-sequence stars over $120 \,\mathrm{d}$. The occurrence frequency of superflares was $2.9\times10^{-3}$ flares per year per star. Therefore, superflares occur more frequently on ZAMS stars than on main-sequence stars.

\section{Discussion}
\label{discussion}

\subsection{Stellar chromospheric activity and starspots of ZAMS stars}

\citet{n15} indicated that the Sun and superflare stars show a rough positive correlation between the amplitude of the light curve and the $r_0$(8542) index, the residual core flux normalized by the continuum level at the line core of the Ca\,\emissiontype{II} IRT. They claimed that the magnetic field strengths of superflare stars are higher than that of the Sun. 

\begin{figure}[htbp]
 \begin{minipage}{0.5\hsize}
  \begin{center}
   \includegraphics[width=8cm]{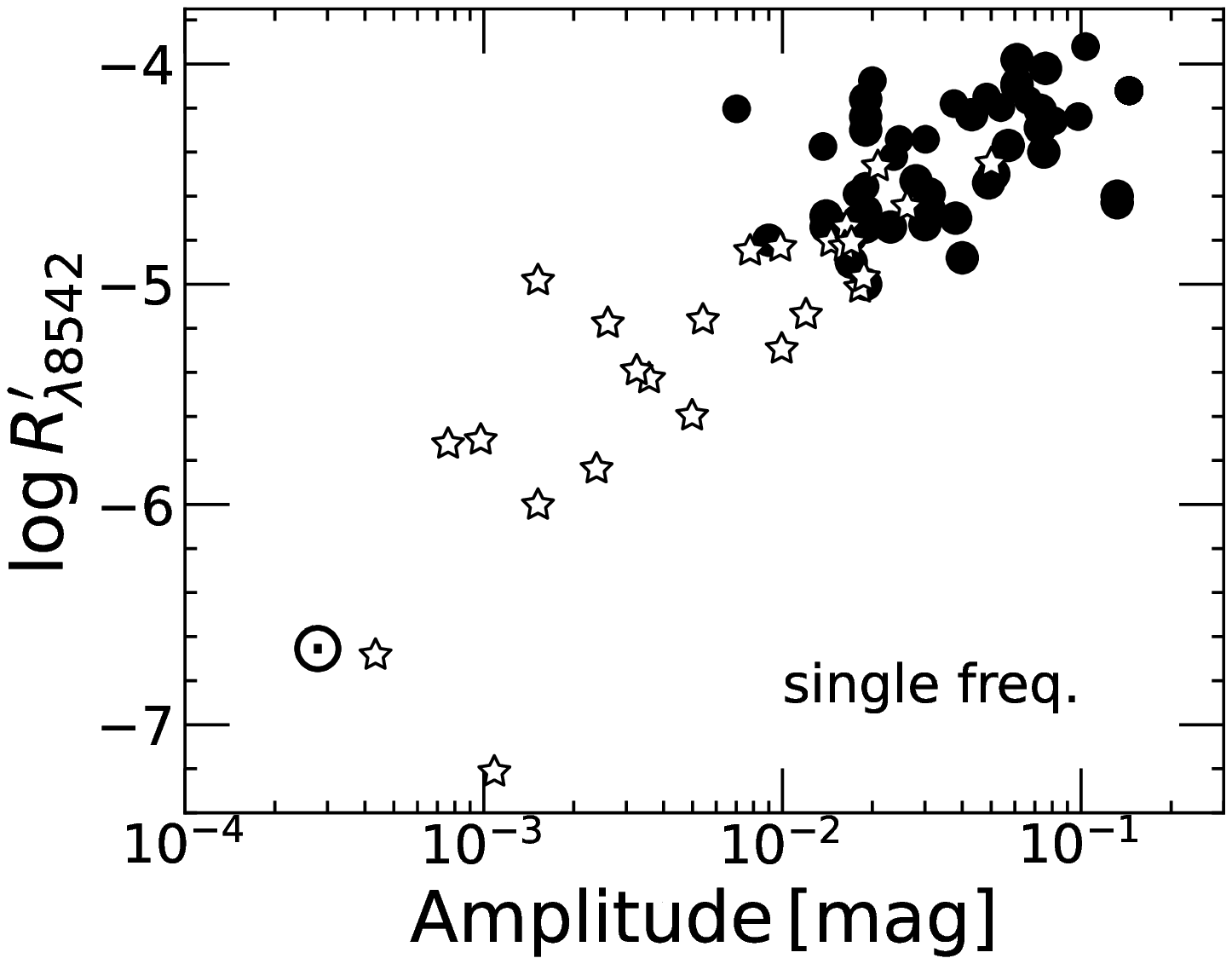}
  \end{center} 
 \end{minipage}
 \begin{minipage}{0.5\hsize}
  \begin{center}
   \includegraphics[width=8cm]{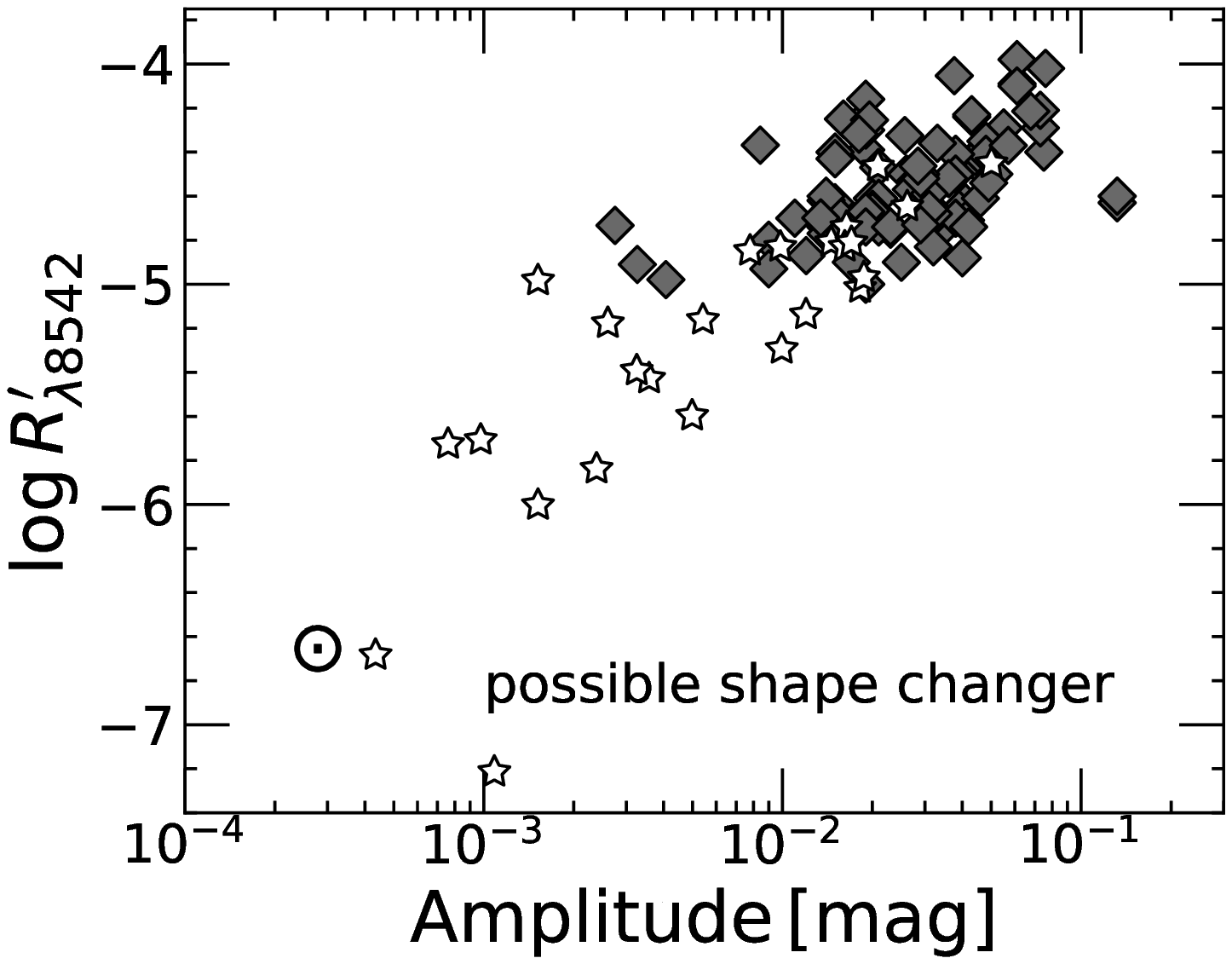}
  \end{center}
 \end{minipage}\\
 \begin{minipage}{0.5\hsize}
  \begin{center}
   \includegraphics[width=8cm]{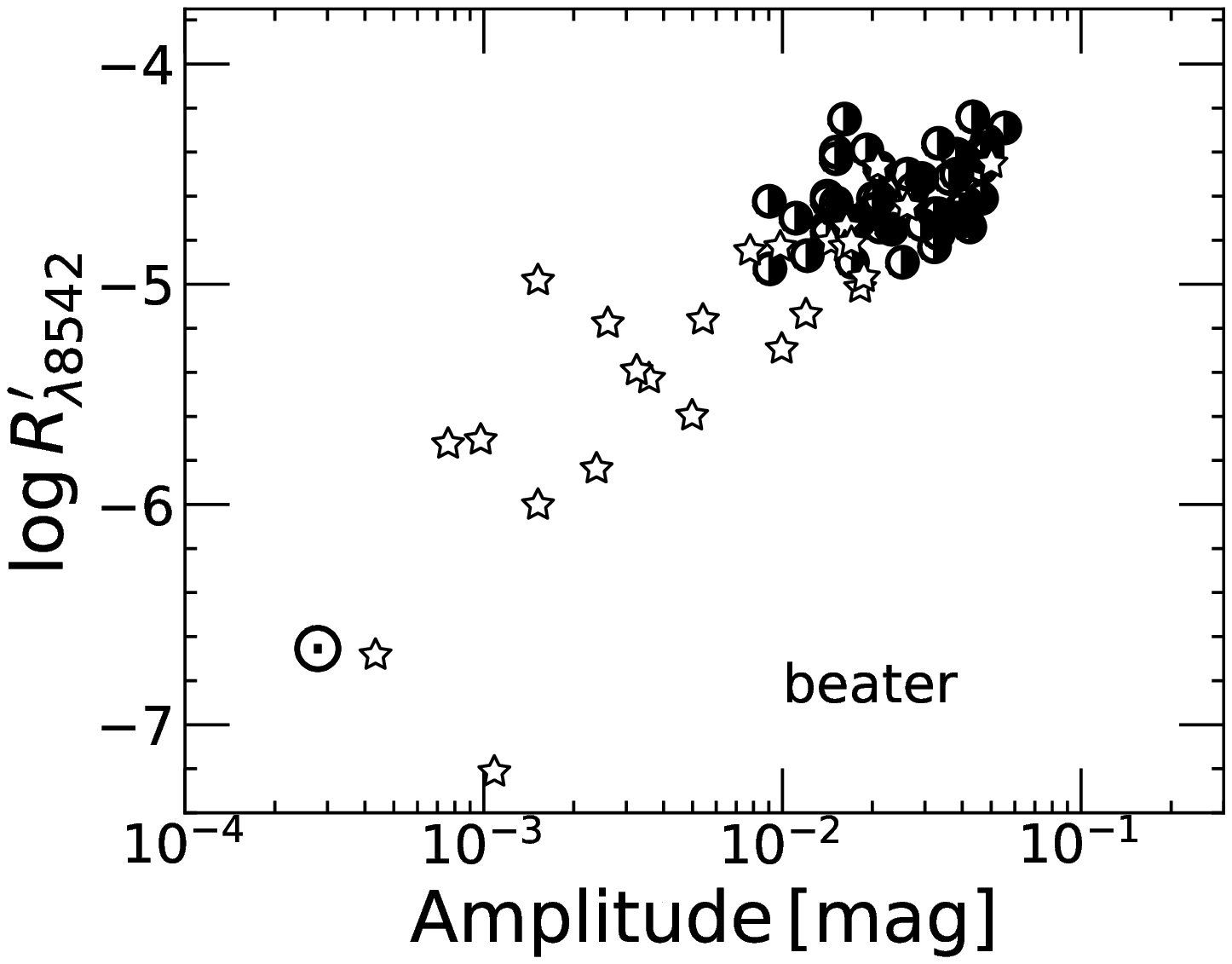}
  \end{center} 
 \end{minipage}
 \begin{minipage}{0.5\hsize}
  \begin{center}
   \includegraphics[width=8cm]{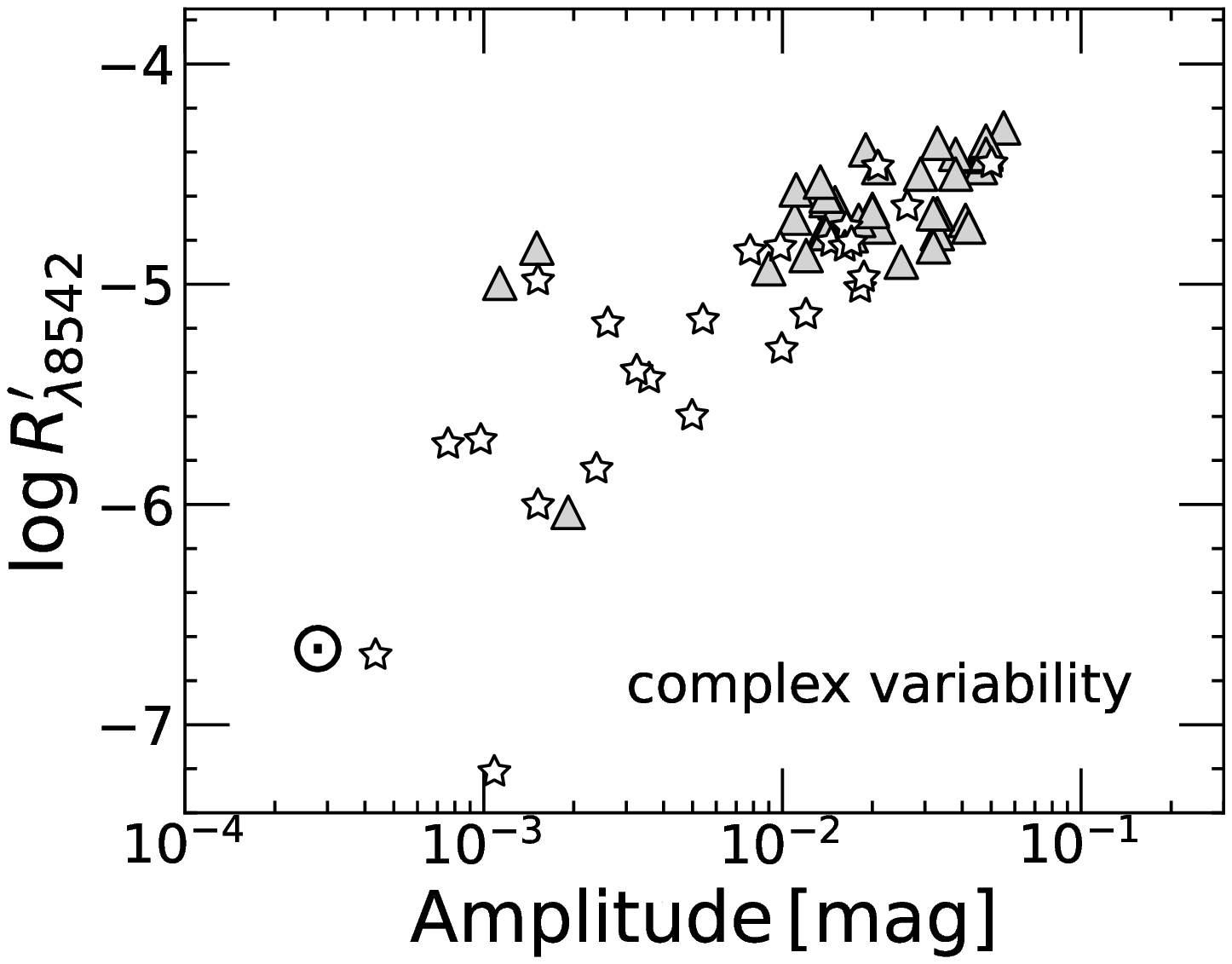}
  \end{center}
 \end{minipage}
 \caption{Relationship between the amplitude of the light curve and the ratio of the surface flux of the Ca\,\emissiontype{II} IRT emission line ($\lambda 8542 \, \mathrm{\AA}$) to the stellar bolometric luminosity, $R^{\prime}_{\lambda 8542}$ of the ZAMS stars in IC 2391, IC 2602, and the Pleiades cluster. The filled circles denote the ZAMS stars with single frequency. The gray diamonds represent the possible shape changer ZAMS stars. The half-filled circles denote the beater ZAMS stars. The light gray triangles show the ZAMS stars with complex variability. Star marks represent superflare stars \citep{n15}. The light variations and $R^{\prime}_{\lambda 8542}$ of the ZAMS stars are as large as those of the most active superflare stars and two orders larger than those of the Sun. The ZAMS stars are located on the extensions of the superflare stars. This suggests that superflare stars link the properties of the Sun to those of the ZAMS stars of ages between $30$ and $120\,\mathrm{Myr}$. }\label{fig:caII2}
\end{figure}

We investigated the relationship between the strength of the Ca\,\emissiontype{II} IRT emission line ($\lambda 8542 \, \mathrm{\AA}$) and the amplitude of the light curve for the ZAMS stars in IC 2391, IC 2602, and the Pleiades cluster. We obtained the amplitude of the light curve of the ZAMS stars in the Pleiades cluster from \citet{r15a}. $R^{\prime}_{\rm \lambda 8542}$ for the ZAMS stars in IC 2391 and IC 2602 was already calculated in Paper I, as listed in \tabref{tab:zams1}. We also obtained $R^{\prime}_{\rm \lambda 8542}$ for the ZAMS stars in the Pleiades cluster from \citet{st97}. We also compare the relationship with the superflare stars studied in \citet{n15}. With $F_{\rm \lambda 8542}$ and $T_{\rm eff}$ listed in Notsu et al. (2015a, b), we calculated $R^{\prime}_{\rm \lambda 8542}$ for the superflare stars (\tabref{tab:sfs}) as
\begin{equation}
	R^{\prime}_{\rm \lambda 8542} = \frac{F_{\rm \lambda 8542} }{\sigma T_{\rm eff}^4}, 
	\label{eq4}
\end{equation}
where $\sigma$ is Stefan-Boltzmann's constant.

\Figref{fig:caII2} shows that the ZAMS stars with a large light curve amplitude have large $R^{\prime}_{\lambda 8542}$. The $R^{\prime}_{\lambda 8542}$ and light curve amplitudes of ZAMS stars are approximately two orders of magnitude larger than those of the Sun. We also found that the ZAMS stars are located on the extensions of the superflare stars \citep{n15}. This result suggests that superflare stars link the properties of the Sun to those of the ZAMS stars of ages between $30$ and $120\,\mathrm{Myr}$. The ZAMS stars belonging in each open cluster are almost evenly distributed. For the ZAMS stars in IC 2391, IC 2602, and the Pleiades cluster, the mean amplitudes are calculated to $0.025\pm0.023\,\mathrm{mag}$, $0.081\pm0.060\,\mathrm{mag}$, and $0.036\pm0.030\,\mathrm{mag}$, respectively. Their mean $R^{\prime}_{\lambda 8542}$ are $10^{-4.79\pm0.22}$, $10^{-4.54\pm0.29}$, and $10^{-4.57\pm0.22}$. The difference of their average values differ less than $1\sigma$. 

Objects with different types of light curves seem to be unevenly distributed in \figref{fig:caII2}, as represented by the colors of the symbols. The ZAMS stars with single frequency have larger light curve amplitudes and larger $R^{\prime}_{\rm \lambda 8542}$ compared with other types, whereas the ZAMS stars with complex variability have smaller light curve amplitudes and smaller $R^{\prime}_{\rm \lambda 8542}$ compared with other types. The possible shape changer ZAMS stars tend to be located between the ZAMS stars with single frequency and the ZAMS stars with complex variability. The relationship between the type of light curve and the magnetic activity will be discussed in the next sub-section.

\begin{figure}[htb]
	\centering
   \includegraphics[width=8cm]{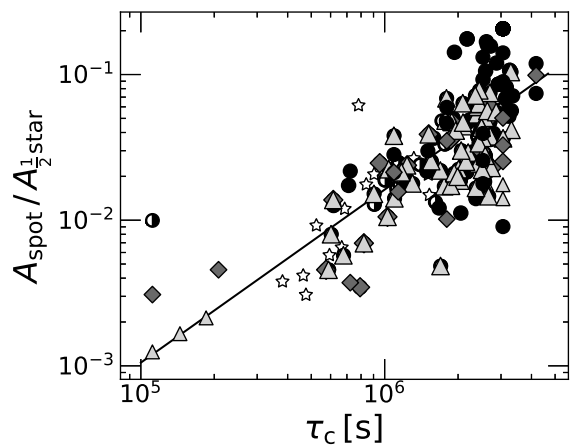} \caption{Coverage of the starspot, $ A_{\rm spot} / A_{\rm \frac{1}{2} star} $, as a function of the convection turnover time, $\tau_{\rm c}$ of the ZAMS stars in IC 2391, IC 2602, and the Pleiades cluster. The filled circles denote the ZAMS stars with single frequency. The gray diamonds show the possible shape changer ZAMS stars. The half-filled circle denotes the beater ZAMS star. The open triangles represent the ZAMS stars with complex variability. The star marks represent superflare stars \citep{n15}. The solid line indicates the least squares linear fit line of all the objects shown here. $ A_{\rm spot} / A_{\rm \frac{1}{2} star} $ shows a positive correlation with $\tau_{\rm c}$. } \label{fig:tau_amp1}
\end{figure}

\Figref{fig:tau_amp1} shows $ A_{\rm spot} / A_{\rm \frac{1}{2} star} $ as a function of the convective turnover time, $\tau_{\rm c}$ of the ZAMS stars in IC 2391, IC 2602, and the Pleiades cluster, and the superflare stars listed in \citet{n15}. In Paper I, we estimated the $\tau_{\rm c}$ values of the ZAMS stars using the $\tau_{\rm c}$ models for stars with the age of $50 \, \mathrm{Myr}$ (IC 2391) and $30 \, \mathrm{Myr}$ (IC 2602) presented in \citet{l10}. We also estimated $\tau_{\rm c}$ for $130 \, \mathrm{Myr}$ (the Pleiades) using the same method. For the ZAMS stars with $T_{\rm eff} > 6180 \, \mathrm{K}$, we applied the approximation of $\tau_{\rm c}$ of \citet{noyes}. We also estimated $\tau_{\rm c}$ for the solar-type superflare stars \citep{n15} by applying the approximation of $\tau_{\rm c}$ of \citet{noyes}, where $(B-V)_0$ is referenced from \citet{re15}. We have estimated $ A_{\rm spot} / A_{\rm \frac{1}{2} star} $ with \equref{eq:spot}, as mentioned above. $T_{\rm eff}$ values of the Pleiades members and the solar-type superflare stars are taken from {\it Gaia} Data Release 2 (DR2; \cite{g18}) and \citet{n15}, respectively. 

\Figref{fig:tau_amp1} indicated that $ A_{\rm spot} / A_{\rm \frac{1}{2} star} $ shows a positive correlation with $\tau_{\rm c}$. The linear fit, $A_{\rm spot} / A_{\rm \frac{1}{2} star} = 10^{-8.7}\cdot \tau_{\rm c}^{1.2} $, was derived. $\tau_{\rm c}$ used in this study is the local convective turnover time. According to the result of \citet{jk07}, there is an approximately linear relationship between $\tau_{\rm c}$ and the global convective turnover time, $\tau_{\rm g}$. For a $1\,\mathrm{M_\odot}$ star at ZAMS, we obtain $\tau_{\rm g} = 1.7 \tau_{\rm c}$. \citet{a16} investigated the relationship between the ratio of the thickness of the convective zone to the stellar radius, $H_{\rm c}/R_*$ and $\tau_{\rm g}$. We derived a linear fit as $\tau_{\rm g} \,\mathrm{[d]} = 10^{3.1}\cdot(H_{\rm c}/R_*)^{2.2}$. 
\begin{equation}
\left\{
\begin{array}{l}
A_{\rm spot} / A_{\rm \frac{1}{2} star} = 10^{-8.7}\cdot \tau_{\rm c}^{1.2} \\
\tau_{\rm g} = 1.7 \tau_{\rm c} \\
\tau_{\rm g} \,\mathrm{[d]} = 10^{3.1}\cdot(H_{\rm c}/R_*)^{2.2}
\end{array}
\right.
\end{equation}
Therefore 
\begin{equation}
\frac{A_{\rm spot} }{ A_{\rm \frac{1}{2} star} } = 4.6 \left(\frac{H_{\rm c}}{R_*}\right)^{2.6}. 
\label{eq:re_sp}
\end{equation}
When we assume that the geometry of a starspot is single circle whose diameter is $D_{\rm spot}$, we can obtain 
\begin{equation}
\left(\frac{D_{\rm spot} }{2R_*}\right)^{2} = 4.6 \left(\frac{H_{\rm c}}{R_*}\right)^{2.6}. 
\end{equation}
Therefore, 
\begin{equation}
D_{\rm spot} = 2 \sqrt{ 4.6 H_{\rm c}{}^{2.6} } \sim 4.3 H_{\rm c}{}^{1.3}. 
\end{equation} 
Numerical simulation indicates that for a large sunspot, the ratio of the horizontal size at the photosphere to $H_{\rm c}$ ranges from $2.04$ to $6.75$ \citep{v71}. \citet{mu73} discussed the relationship between the thickness of the convection zone and the size of a starspot. They considered that the depth of the spot, $H_{\rm s}$, is equal to the depth of the convection zone, $H_{\rm c}$, which yields $2.04 < D_{\rm spot}/H_{\rm c} < 6.75$. We derived $D_{\rm spot}/H_{\rm c} \sim 4.3$, which is well within the values presented in \citet{mu73}. 

\begin{table}[htb]
\caption{Rossby numbers and $R^{\prime}$ of solar-type superflare stars listed in \citet{n15}} \label{tab:sfs}
\centering
\begin{tabular}{lcc}
\hline
Object name        & $\log\, N_{\rm R}$    & $\log\,R^{\prime}_{\lambda 8542}$ \\ 
(1) & (2) & (3) \\ \hline
KIC 4742436  & -0.48 & -5.60 \\
KIC 6865484  & 0.23  & -4.85 \\
KIC 7354508  & 0.34  & -5.16 \\
KIC 7420545  & 0.31  & -5.14 \\
KIC 8802340  & -0.13 & -5.02 \\
KIC 9583493  & -0.25 & -4.81 \\
KIC 9652680  & -0.78 & -4.45 \\
KIC 10471412 & 0.44  & -5.18 \\
KIC 10528093 & -0.10 & -4.47 \\
KIC 11140181 & 0.00  & -4.74 \\
KIC 11303472 & -0.11 & -4.83 \\
KIC 11390058 & 0.35  & -5.43 \\
KIC 11455711 & 0.24  & -5.29 \\
KIC 11494048 & 0.53  & -5.39 \\
KIC 11764567 & 0.22  & -4.97 \\
KIC 12266582 & -0.18 & -4.81 \\ \hline
\end{tabular}
\end{table}

\subsection{Starspots and Rotation-Activity Relationship} 

\begin{figure}[htb]
	\centering
	\includegraphics[height=70mm]{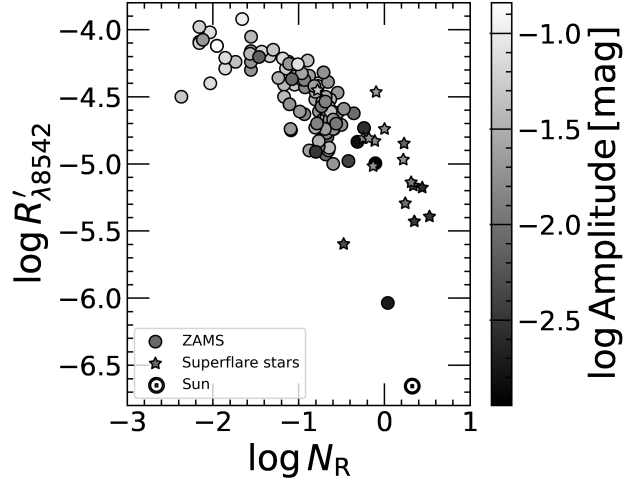}
	\caption{Relationship between the ratio of the surface flux of the Ca\, \emissiontype{II} emission line ($\lambda 8542 \, \mathrm{\AA}$) to the stellar bolometric luminosity, $R^{\prime}_{\lambda 8542}$, and the Rossby number, $N_{\rm R}$, of the ZAMS stars in IC 2391, IC 2602, and the Pleiades cluster (circles). Star marks represent superflare stars \citep{n15}. The gray scale of all the symbols represents the amplitude of the light curve. Objects with a smaller $N_{\rm R}$ have both a saturated $R^{\prime}_{\lambda 8542}$ and a larger light curve amplitude. } \label{fig:NvsR2}
\end{figure}

In this section, we examine the relationship between the spot area and the rotation-activity relationship. In Paper I, we calculated $N_{\rm R}$ for the ZAMS stars in IC 2391 and IC 2602. We also calculated $N_{\rm R}$ for the Pleiades members and the superflare stars listed in \citet{n15}. Their $P$ values were taken from \citet{r15b} and \citet{n15}, respectively. $N_{\rm R}$ for the superflare stars is shown in \tabref{tab:sfs}. 
The relationship between $N_{\rm R}$ and the ratio of the surface flux of the Ca\,\emissiontype{II} IRT emission line to the stellar bolometric luminosity, $R^{\prime}_{\rm \lambda 8542}$, is shown in \figref{fig:NvsR2}. For the ZAMS stars in IC 2391, IC 2602, and the Pleiades cluster with $N_{\rm R} > 10^{-1.1}$, $R^{\prime}_{\lambda 8542}$ increases with decreasing $N_{\rm R}$ until it saturates. For the ZAMS stars with $N_{\rm R} < 10^{-1.1}$, $R^{\prime}_{\lambda 8542}$ reaches a constant level. All the superflare stars and the Sun belong to the unsaturated regime. The ZAMS stars belonging in each open cluster are almost evenly distributed. 

Objects with a smaller $N_{\rm R}$ have both a saturated $R^{\prime}_{\lambda 8542}$ and a larger light curve amplitude (\figref{fig:NvsR2}). It is known that small $N_{\rm R}$ objects are strongly activated by their dynamo process. The dynamo process is considered to induce a strong magnetic field, resulting in large starspots and saturated strong emission lines. The Ca\, \emissiontype{II} saturation is consistent with the existence of a large-scale starspot or starspot group. 

\begin{figure}[htbp]
 \begin{minipage}{0.5\hsize}
  \begin{center}
   \includegraphics[height=70mm]{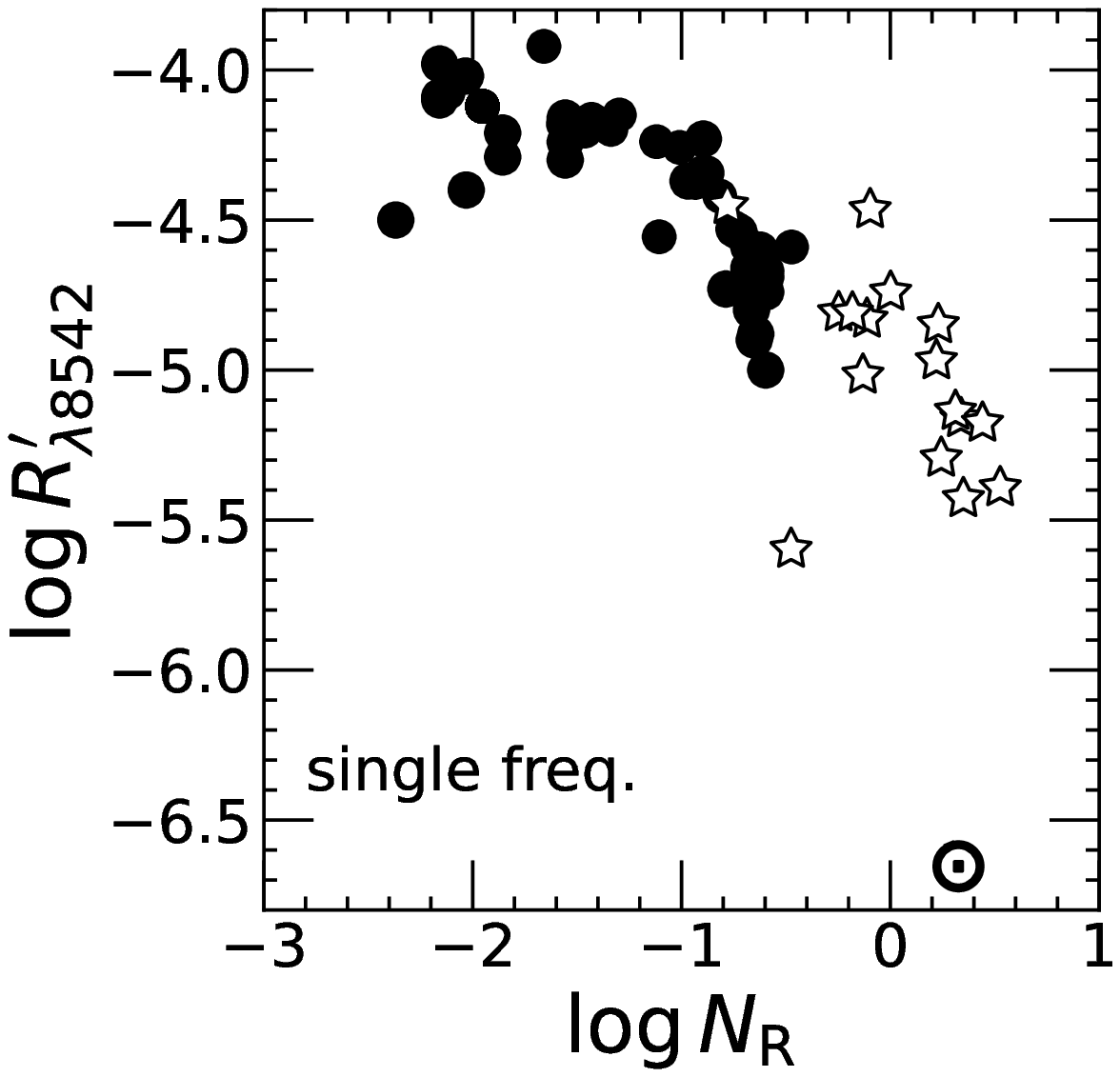}
  \end{center} 
 \end{minipage}
 \begin{minipage}{0.5\hsize}
  \begin{center}
   \includegraphics[height=70mm]{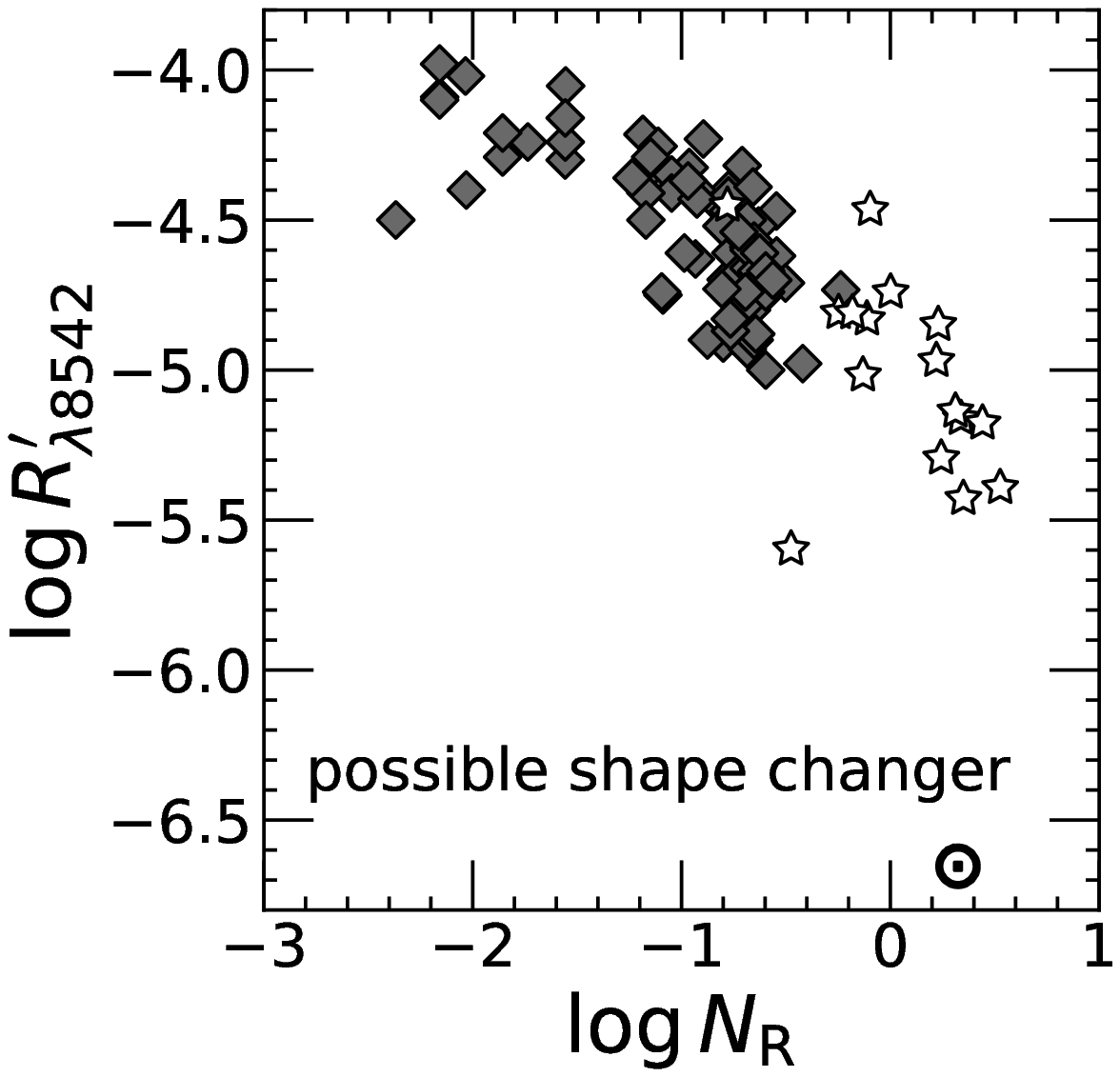}
  \end{center}
 \end{minipage}\\
 \begin{minipage}{0.5\hsize}
  \begin{center}
   \includegraphics[height=70mm]{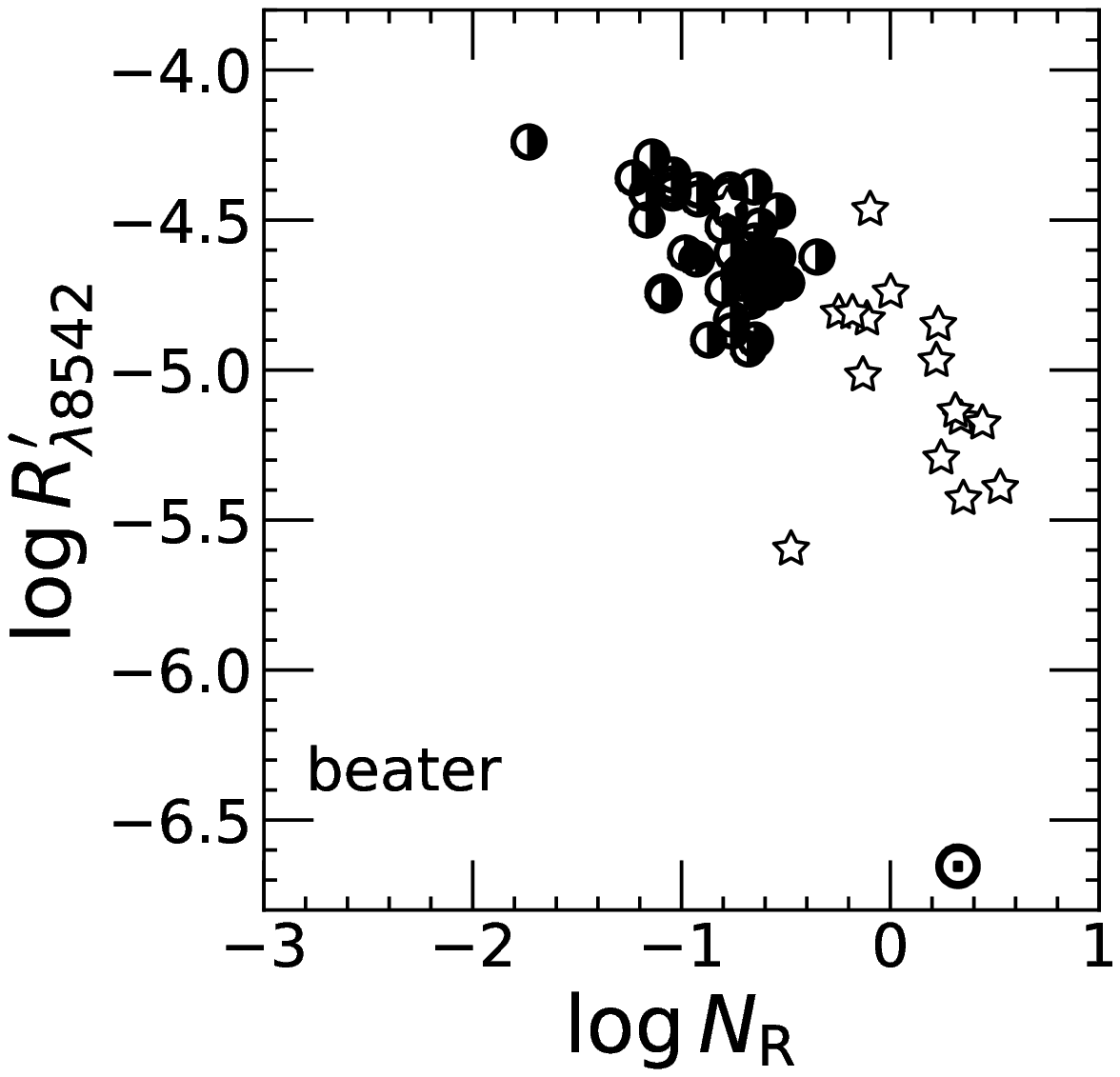}
  \end{center} 
 \end{minipage}
 \begin{minipage}{0.5\hsize}
  \begin{center}
   \includegraphics[height=70mm]{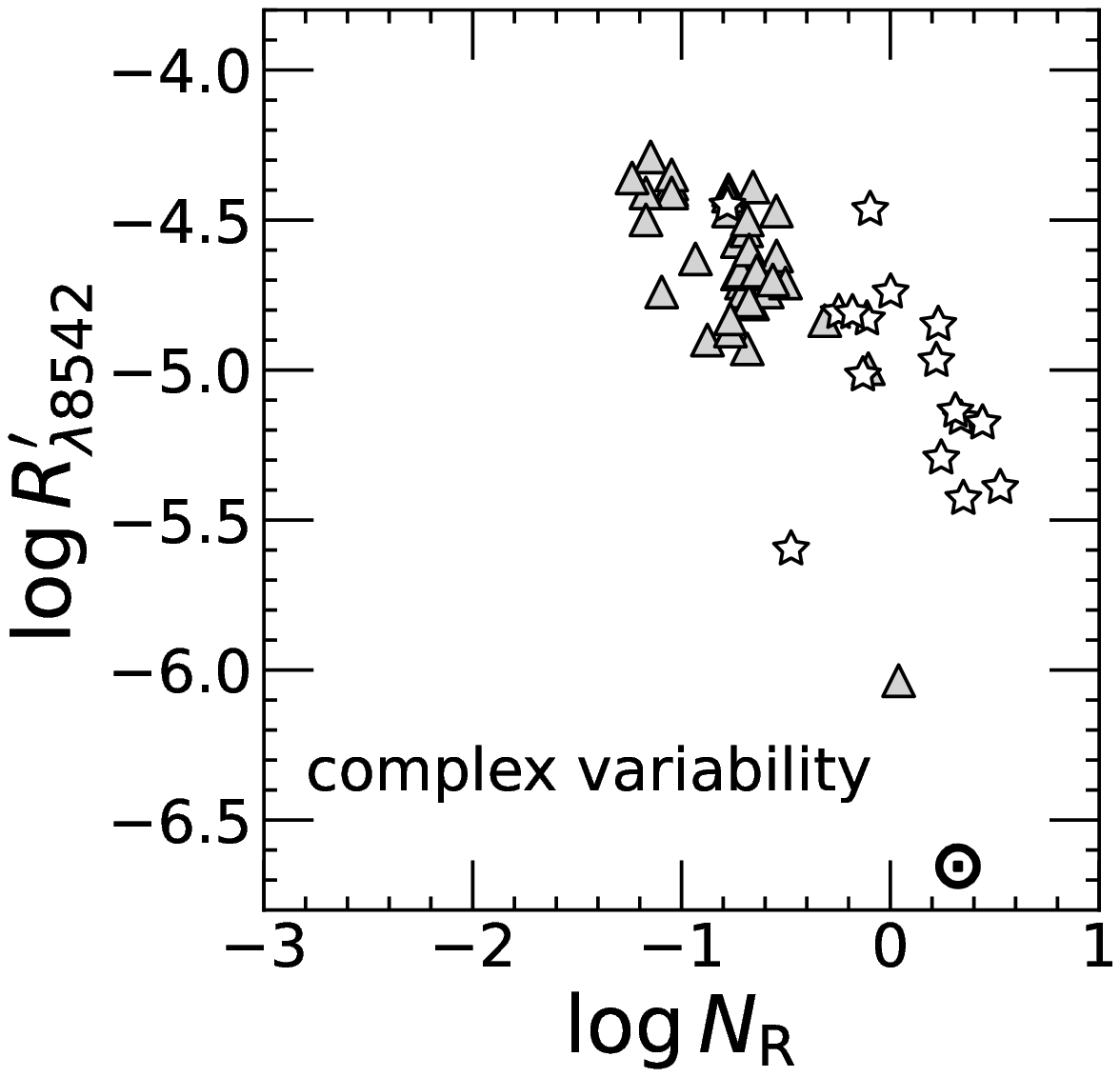}
  \end{center}
 \end{minipage}
 \caption{Relationship between the ratio of the surface flux of the Ca\,\emissiontype{II} emission line ($\lambda 8542 \, \mathrm{\AA}$) to the stellar bolometric luminosity, $R^{\prime}_{\lambda 8542}$, and the Rossby number, $N_{\rm R}$, of the ZAMS stars in IC 2391, IC 2602, and the Pleiades cluster. 
The filled circles denote ZAMS stars with single frequency. The gray diamonds denote the possible shape changer ZAMS stars. The half-filled circles denote the beater ZAMS stars. The open triangles denote ZAMS stars with complex variability. Star marks represent superflare stars \citep{n15}. The ZAMS stars with single frequency or possible shape change have small $N_{\rm R}$, large $R^{\prime}_{\lambda 8542}$, and larger light curve amplitude, while the ZAMS stars with beat or complex variability have large $N_{\rm R}$, small $R^{\prime}_{\lambda 8542}$, and smaller light curve amplitude. } \label{fig:NvsR3}
\end{figure}

\Figref{fig:NvsR3} shows the relationship between $R^{\prime}_{\lambda 8542}$ and $N_{\rm R}$ of the ZAMS stars in IC 2391, IC 2602, and the Pleiades cluster. We noticed that the ZAMS stars with single frequency or possible shape change have small $N_{\rm R}$ and large $R^{\prime}_{\lambda 8542}$. By combining with \figref{fig:NvsR2}, it is suggested that those objects show larger light curve amplitude, indicating huge spot/spot groups. The ZAMS stars with beat or complex variability have large $N_{\rm R}$, small $R^{\prime}_{\lambda 8542}$, and smaller light curve amplitude. This result suggests that those ZAMS stars do not have huge spot/spot groups. 

We consider that the strong magnetic field generates a limited number of large starspots. With \textit{Kepler} observations, \citet{re20} revealed that $369$ stars with near-solar rotational periods have much stronger and more regular light variation than the Sun. \citet{i20} constructed the light curve model of such stars by injecting active regions on stellar surfaces with highly inhomogeneous distribution compared with the Sun. The injected active region consists of spot groups and faculae. They used wavelength-dependent contrast models of faculae and spots on the Sun calculated by \citet{u99}. Different activity levels lead to different types of light curves, i.e., periodic, strong variation, and complex variation. They qualified the activity level in terms of the chromospheric S-index. By extrapolating the relationship between the disk coverage by spots and the S-index of the Sun \citep{s14}, they obtained the S-index for a more active star with a given light curve. \citet{i20} found that the light curve amplitude increases with larger S-index because of the increasing occurrence of active regions. Active longitude nesting, in which active regions appear on the opposite side, leads to highly regular variability. The regularity and light curve amplitude decrease for small S-index objects. As mentioned above, $R^{\prime}_{\lambda 8542}$ of the ZAMS stars in IC 2391, IC 2602, and the Pleiades cluster with single frequency tend to be saturated, and that of the ZAMS stars with complex variability is relatively small. This result supports the assumption in \citet{i20} that the active regions are inhomogeneously distributed on the surface of an active star. 

\subsection{Period-Color Distribution}
\citet{r20} investigated the relationship between $(V-K_{\rm s})_0$ and $P$ with $K2$ data for six clusters: Rho Oph, Taurus, Upper Sco, Taurus foreground, Pleiades, and Prasepe ($1-790\,\mathrm{Myr}$) and discussed spin-down evolution to the main-sequence. 

\begin{figure}[htb]
	\centering
   \includegraphics[width=8cm]{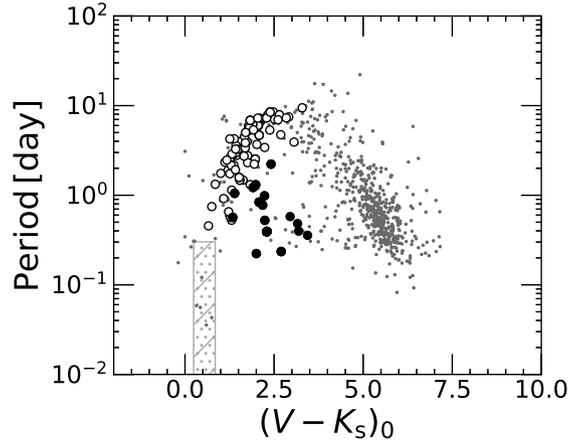} \caption{Relationship between the $(V-K_{\rm s})_0$ color and the rotational period of ZAMS stars in IC 2391, IC 2602, and the Pleiades cluster. The filled circles show the ZAMS stars whose Ca\,\emissiontype{II} IRT emission lines are saturated ($R^{\prime}_{\lambda 8542} \sim 10^{-4.2}$). The open circles show the ZAMS stars with $R^{\prime}_{\lambda 8542} < 10^{-4.2}$ whose Ca\,\emissiontype{II} IRT emission lines are unsaturated. The tiny circles denote the Pleiades members examined by \citet{r15a} but whose Ca\,\emissiontype{II} IRT emission lines had not been observed in \citet{st97}. The hatched area represents $0.24 < (V-K_{\rm s})_0 < 0.85$ and $P \leq 0.3 \, \mathrm{days}$. The ZAMS stars whose Ca\, \emissiontype{II} emission lines are saturated tend to be distributed in the lower part of the figure, while the ZAMS stars whose Ca\, \emissiontype{II} emission lines are unsaturated are located close to the majority of the Pleiades members. }\label{fig:pul} 
\end{figure}

\begin{figure}[htbp]
 \begin{minipage}{0.5\hsize}
  \begin{center}
   \includegraphics[width=8cm]{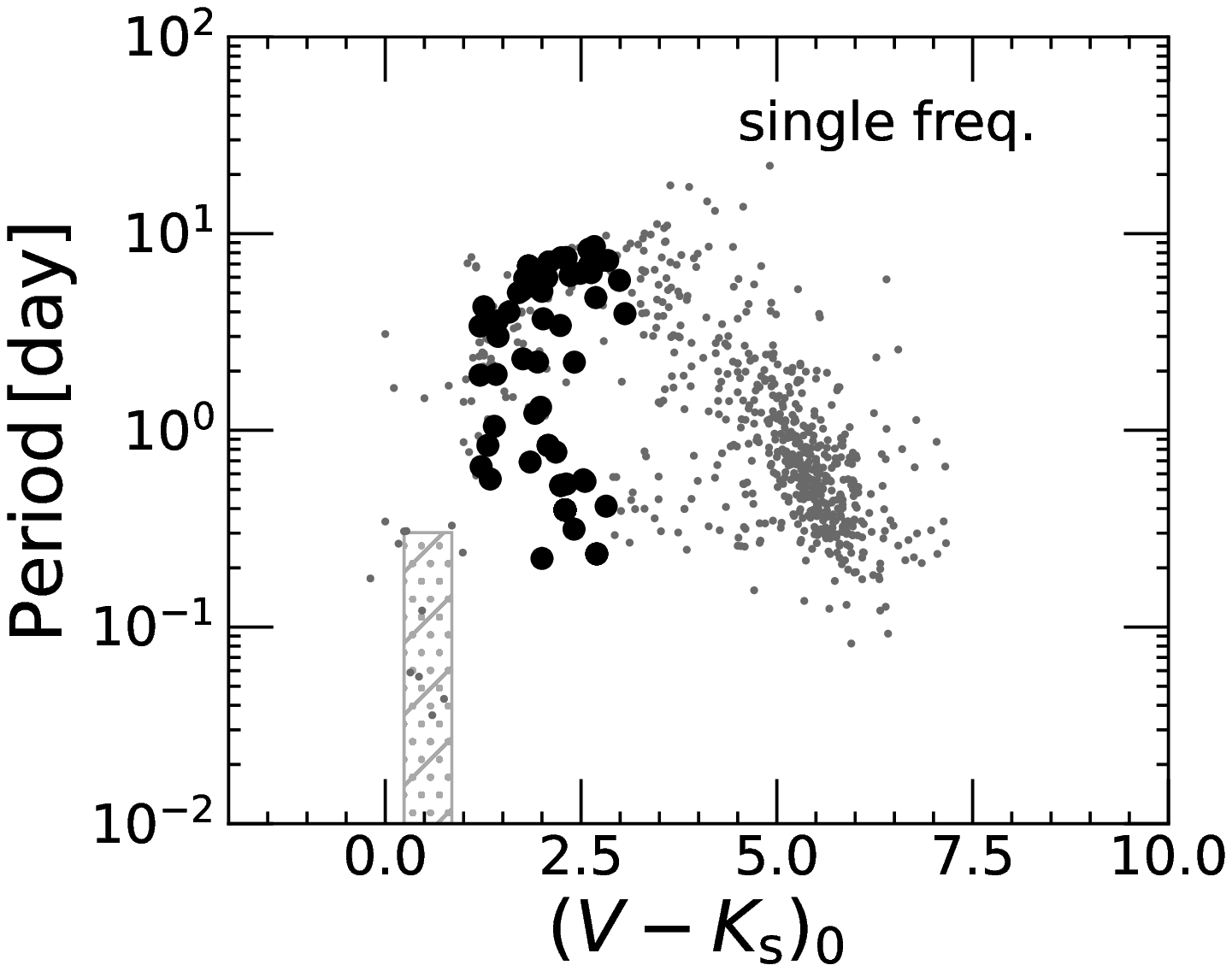}
  \end{center} 
 \end{minipage}
 \begin{minipage}{0.5\hsize}
  \begin{center}
   \includegraphics[width=8cm]{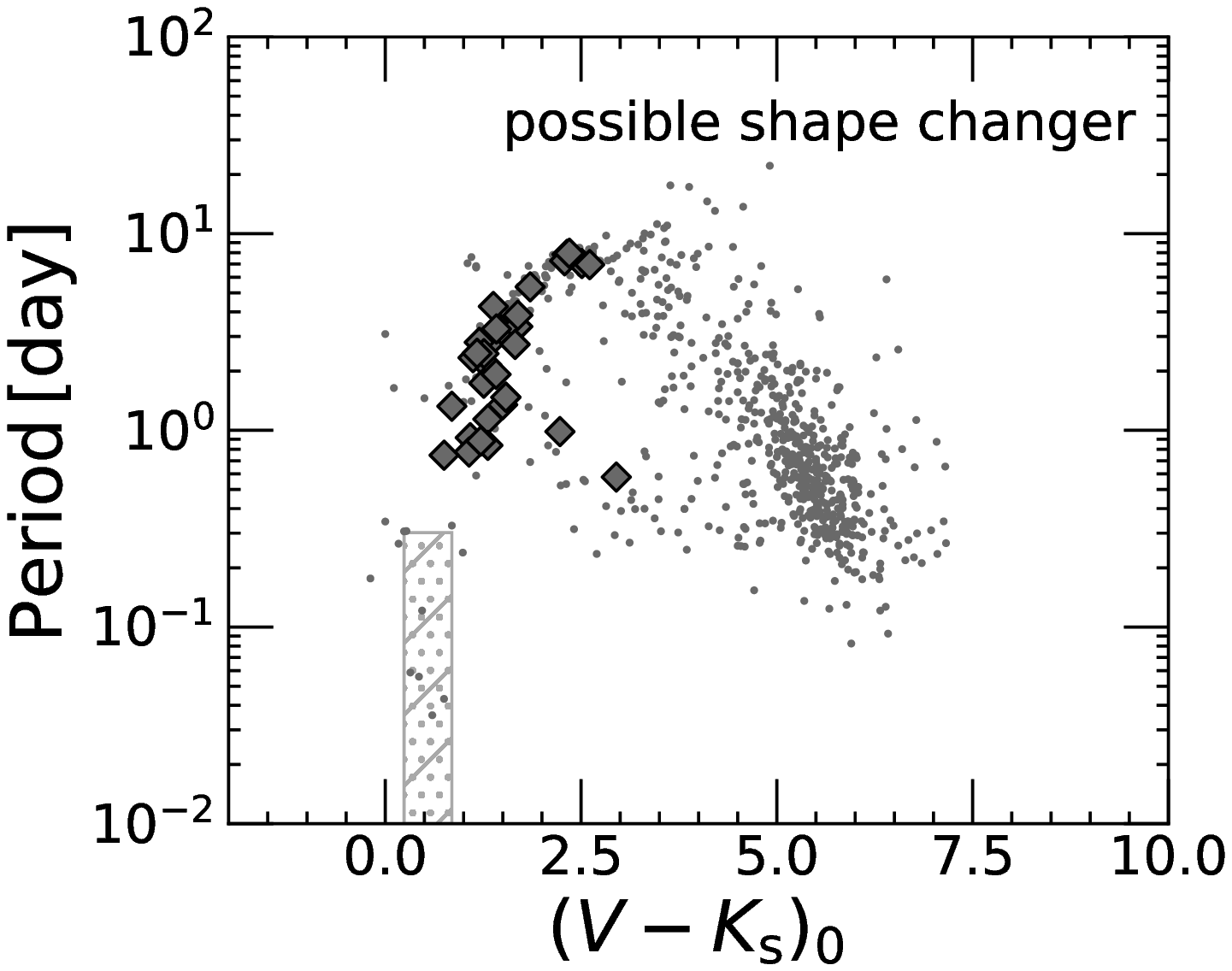}
  \end{center}
 \end{minipage}\\
 \begin{minipage}{0.5\hsize}
  \begin{center}
   \includegraphics[width=8cm]{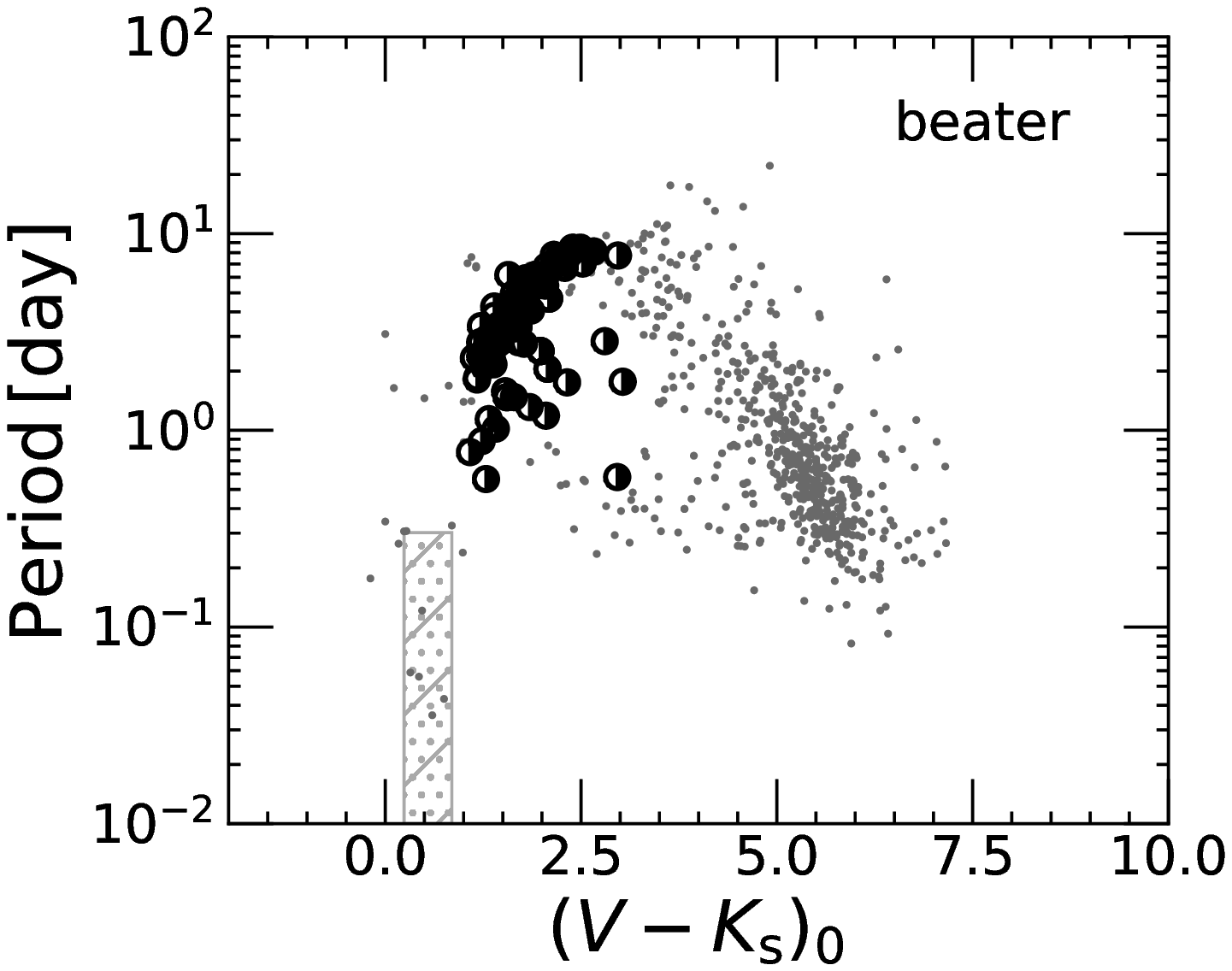}
  \end{center} 
 \end{minipage}
 \begin{minipage}{0.5\hsize}
  \begin{center}
   \includegraphics[width=8cm]{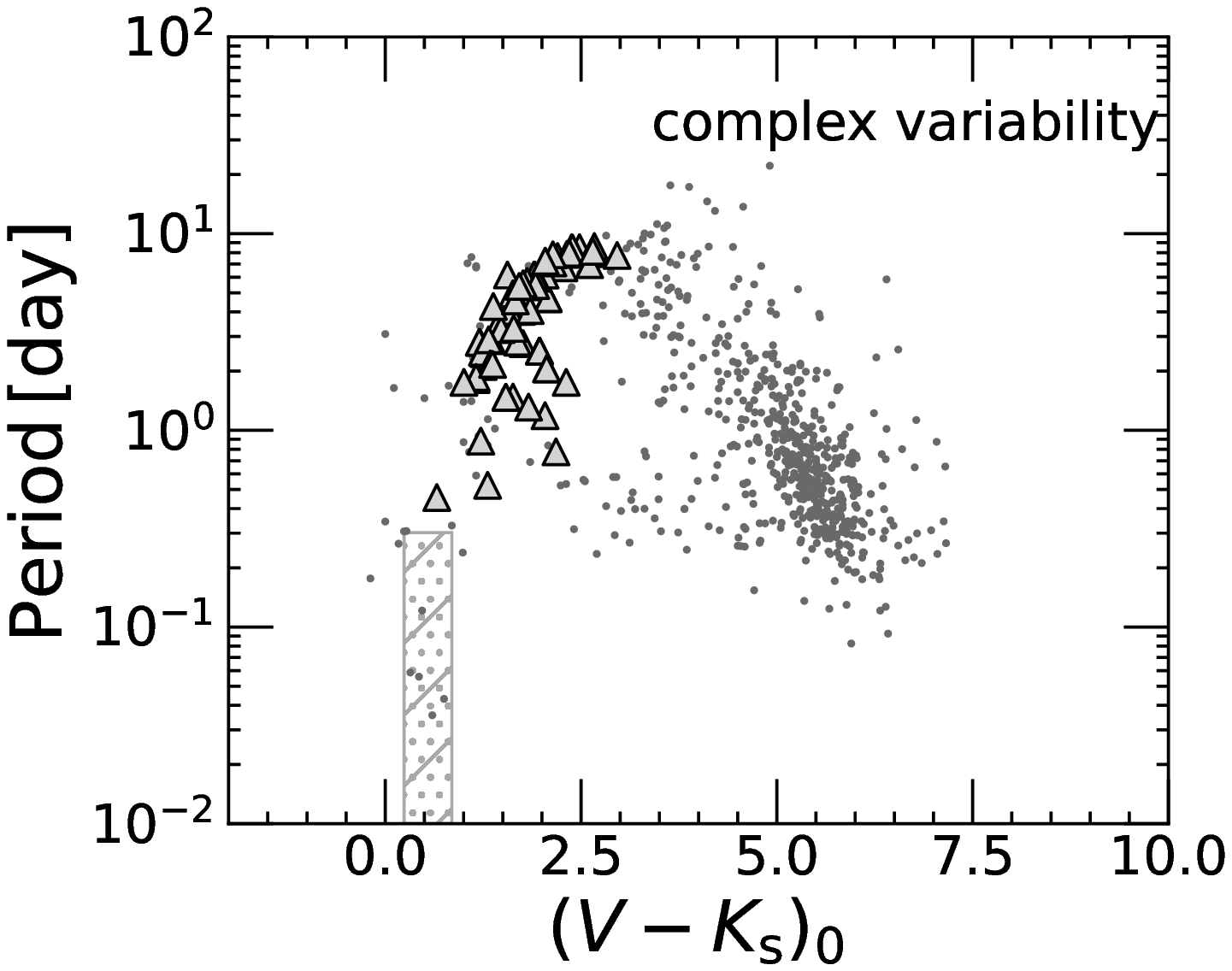}
  \end{center}
 \end{minipage}
 \caption{Relationship between the $(V-K_{\rm s})_0$ color and the rotational period of ZAMS stars in IC 2391, IC 2602, and the Pleiades cluster. The filled circles denote the ZAMS stars with single frequency. The gray diamonds denote the possible shape changer ZAMS stars. The half-filled circles denote the beater ZAMS stars. The open triangles denote the ZAMS stars with complex variability. The tiny circles denote the Pleiades members examined by \citet{r15a} but whose Ca\,\emissiontype{II} IRT emission lines had not been observed in \citet{st97}. The hatched area represents $0.24 < (V-K_{\rm s})_0 < 0.85$ and $P \leq 0.3 \, \mathrm{days}$. Most of the ZAMS stars rotating rapidly have light curve with a single frequency period. } \label{fig:pul2} 
\end{figure}

We investigated the $P$ of the members of IC 2391 and IC 2602 as a function of $(V-K_{\rm s})_0$ (\figref{fig:pul} and \figref{fig:pul2}). We took the $V$ magnitudes from \citet{m09} and the $K_{\rm s}$ magnitudes from the 2MASS survey \citep{c03}. These values are listed in \tabref{tab:zams1}. The $(V - K_{\rm s})$ values were converted to unreddened values, $(V - K_{\rm s})_0$, using the reddening parameters for the clusters; $E(B - V) = 0.006$ for IC 2391 from \citet{ps96} and $E(B - V) = 0.04$ for IC 2602 from \citet{b61} were converted to $E(V - K_{\rm s}) = 0.016$ and $E(V - K_{\rm s}) = 0.110$ for IC 2391 and IC 2602 by multiplying the $E(V - K_{\rm s}) / E(B - V)$ value given in \citet{r85}. 

We found that the ZAMS stars in IC 2391, IC 2602, and the Pleiades cluster whose Ca\, \emissiontype{II} emission lines are saturated tend to be distributed in the lower part of the \figref{fig:pul}, while the ZAMS stars whose Ca\, \emissiontype{II} emission lines are unsaturated are located close to the majority of the Pleiades members. It is possible that the ZAMS stars whose Ca\, \emissiontype{II} emission lines are saturated have not reached the spin down stage yet. In \figref{fig:pul2}, we also found that most of the ZAMS stars rotating rapidly have a single frequency period. These results obtained from figures 7 and 8 show that the ZAMS stars not reached the spin down stage have saturated Ca\, \emissiontype{II} emission lines and single frequency periods. As mentioned in the discussion on \figref{fig:NvsR3}, the ZAMS stars with single frequency tend to have strong Ca\,\emissiontype{II} IRT emission lines. These results are consistent with \citet{r15b} and \citet{st16}, in which approximately half of the ZAMS stars rotating rapidly have sinusoidal light curves. Here, we do not think that stellar pulsation is the cause of the light variation. \citet{r15a} considered that only $1\%$ of the Pleiades members are pulsators, including the $\delta$ Sct-type, which have very small light curve amplitudes, $0.24 < (V-K_{\rm s})_0 < 0.85$, and short rotational periods. No ZAMS stars in IC 2391 and IC 2602 have both a short period and an early spectral type.

\subsection{Superflare} \label{flare}

\begin{figure}[htb]
	\centering
	\includegraphics[height=60mm]{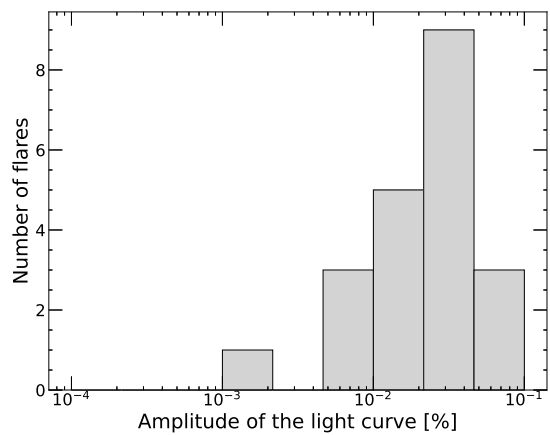}
	\caption{Relationship between the light curve amplitude and the number of flares on the ZAMS stars in IC 2391 and IC 2602. More frequent flares are observed on the ZAMS stars with larger light curve amplitudes. } \label{fig:flare1}
\end{figure}

After removing the rotational light variations in the {\it TESS} light curves, we detected $21$ flares from $12$ ZAMS stars in IC 2391 and IC 2602. These flares show characteristics similar to those of solar-type main-sequence stars. \figref{fig:flare1} shows the relationship between the light curve amplitude and the number of the flares on the ZAMS stars in IC 2391 and IC 2602. More frequent flares are observed on the ZAMS stars with larger light curve amplitudes. We found no correlation between the rotational phase and the number of flares, which is also claimed in a study on M dwarfs \citep{d18}. 

\begin{figure}[htb]
	\centering
	\includegraphics[width=7cm]{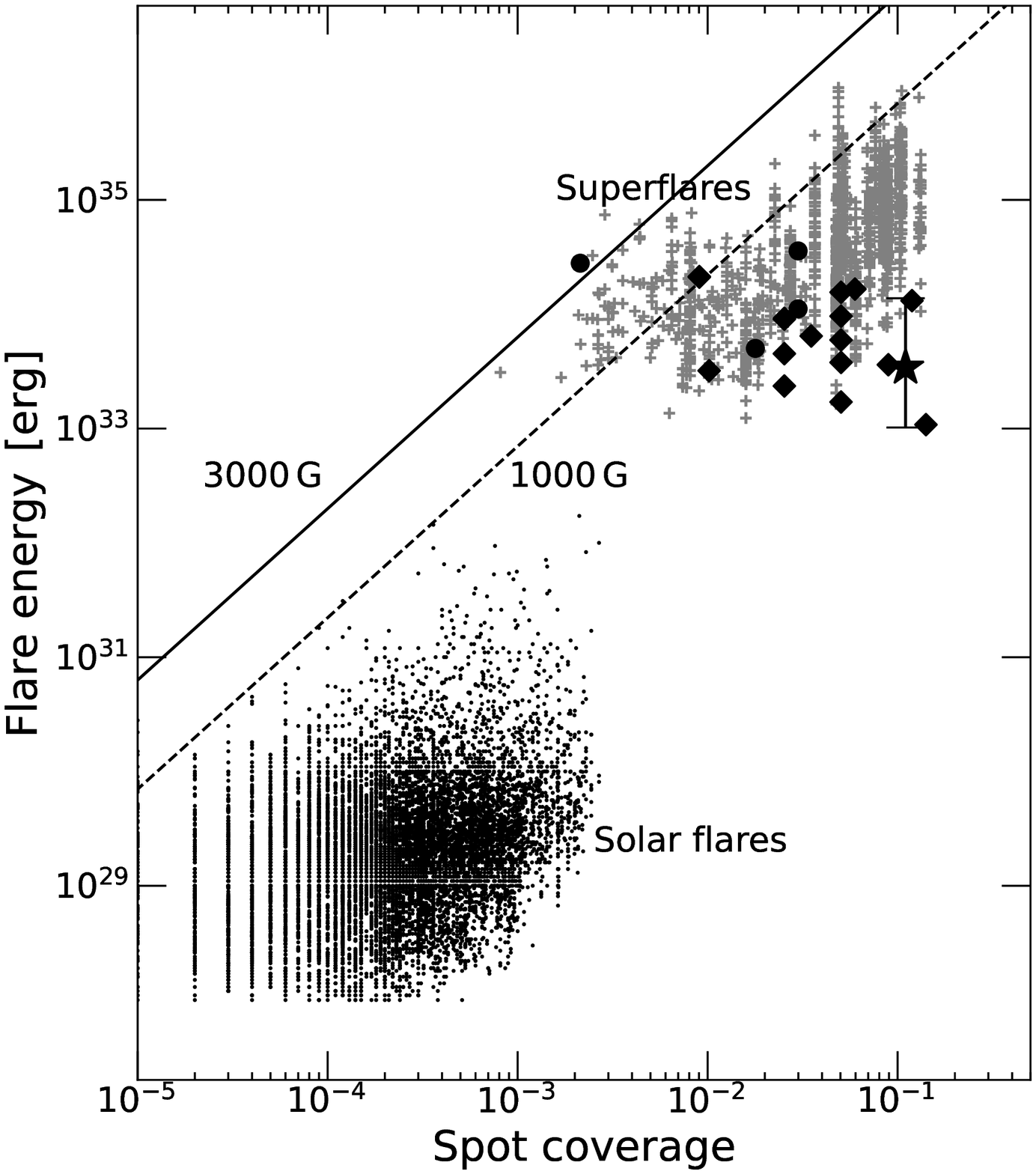}
	\caption{Relationship between spot coverage, $A_{\rm spot} / A_{\rm \frac{1}{2} star}$, and flare energy, $E_{\rm flare}$. Circles and diamonds denote flares on the ZAMS stars in IC 2391 and IC 2602, respectively. The star symbol represents the flare distribution of EK Dra detected by \textit{TESS} \citep{na21}. The tiny circles show solar flares \citep{s13}, and the cross symbols show superflares on Sun-like stars with $T_{\rm eff}=5100-5600\,\mathrm{K}$. The black solid and dashed lines correspond to the analytical relationship between $E_{\rm flare}$ and $A_{\rm spot}$ for $B = 3000\,\mathrm{G}$ and $1000 \,\mathrm{G}$, respectively. For the flares on the ZAMS stars in IC 2391 and IC 2602, the energy is estimated to be $\sim 10^{33}-10^{35}\,\mathrm{erg}$, which is comparable with the energy of a superflare. } \label{fig:flare2}
\end{figure}

The energy of a flare can be estimated from the amplitude of a flare ($\Delta F_{\rm flare}/F$) and the duration time. A detailed description is presented in \citet{s13} and their related studies. $\Delta F_{\rm flare}/F$ can be written as
\begin{equation}
	\frac{\Delta F_{\rm flare}}{F} = \frac{ A_{\rm flare} }{ A_{\rm \frac{1}{2} star} } \frac{\int B(T_{\rm flare}, \lambda) S(\lambda) d\lambda }{ \int B(T_{\rm eff}, \lambda) S(\lambda) d\lambda } , 
	\label{eq3}
\end{equation}
where $A_{\rm flare} / A_{\rm \frac{1}{2} star}$ is the fraction of the flare-emitting area normalized by the effective area of the stellar hemisphere. With the Stefan-Boltzman law regarding the photospheric luminosity and $T_{\rm eff}$ of the ZAMS stars in {\it Gaia} DR2, we estimated the stellar radius and converted it to $A_{\rm \frac{1}{2} star}$. $T_{\rm flare}$ is the effective temperature of a flare component. We assumed that the spectral energy distribution of the flare component is similar to blackbody radiation, with an effective temperature of $10000\,\mathrm{K}$ (e.g., Hawley \& Pettersen 1991; Hawley \& Fisher 1992). $B(T_{\rm flare}, \lambda)$ is the Plank function, and $S(\lambda)$ is the spectral response function of the TESS detector ($\lambda 6000-10000\,\mathrm{\AA}$; Ricker et al. 2015). We estimated $A_{\rm flare}$ with \equref{eq3} and substituted it into \equref{eq6}, after which we obtained the bolometric luminosity of the flare, $L_{\rm flare}$. 
\begin{equation}
	L_{\rm flare} = \sigma T_{\rm flare}^4 A_{\rm flare}. 
	\label{eq6}
\end{equation}
The total bolometric energy of a superflare, $E_{\rm flare}$, is an integral of $L_{\rm flare}$ over the flare duration, 
\begin{equation}
	E_{\rm flare} = \int L_{\rm flare}(t) dt .
	\label{eq5}
\end{equation}

\Figref{fig:flare2} is a scatter plot of the flare energy, $E_{\rm flare}$, as a function of spot coverage, $A_{\rm spot} / A_{\rm \frac{1}{2} star}$. In the figure, solar flares and superflares, including those of EK Draconis (EK Dra), are also shown. EK Dra is known as an active G-type ZAMS star that exhibits frequent flares. For EK Dra, we referenced the spot coverage from \citet{str98}, and the flare energy from \citet{na21}. We found that the spot coverage of the ZAMS stars in IC 2391 and IC 2602 is similar to that of superflare stars. The energy of the flares is estimated to be $\sim 10^{33}-10^{35}\,\mathrm{erg}$, which is comparable with the energy of a superflare. One flare is above the analytical relationship for $B = 3000\,\mathrm{G}$. 

\begin{figure}[htb]
	\centering
	\includegraphics[width=7cm]{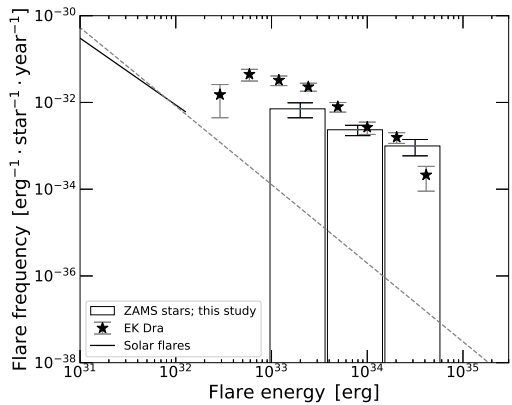}
	\caption{Occurrence rate of flares on ZAMS stars and of solar flares. The histogram shows the frequency distribution of flares on the $12$ ZAMS stars in IC 2391 and IC 2602 on which flares were detected. The star symbols represent the flare distribution of EK Dra detected by \textit{TESS} \citep{na21}. The solid line indicates the power-law distribution of solar flares observed in hard X-rays \citep{c93}, with the distribution $dN/dE \propto E^{-\alpha}$ with index $\alpha \sim 1.8$. The dashed line corresponds to the power-law distribution estimated from superflares in Sun-like stars and solar flares \citep{s13}. The flare frequency on the $12$ ZAMS stars is similar to that of EK Dra. } \label{fig:flare3} 
\end{figure} 

We compare the occurrence rate of flares on ZAMS stars in IC 2391 and IC 2602 with those of solar flares (\figref{fig:flare3}). Flares with larger energy have a lower occurrence rate, the same as solar flares. For the $12$ ZAMS stars from which flares were detected, we obtained the distribution $dN/dE \propto E^{-\alpha}$ with index $\alpha \sim 0.7$. We found that the flare frequency on EK Dra is similar to that of our ZAMS stars. 

\begin{figure}[htb]
	\centering
	\includegraphics[width=7cm]{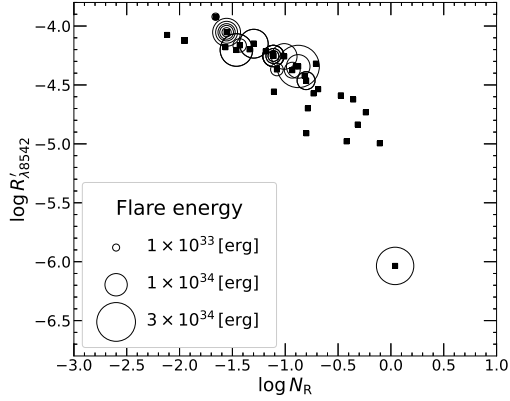}
	\caption{The $21$ flares are shown for the relationship between the ratio of the surface flux of the Ca\, \emissiontype{II} emission line ($\lambda 8542 \, \mathrm{\AA}$) to the stellar bolometric luminosity, $R^{\prime}_{\lambda 8542}$, and the Rossby number, $N_{\rm R}$, of the ZAMS stars in IC 2391 and IC 2602 (squares). The flares and their energies are represented by open circles and their size, respectively. Most of the ZAMS stars on which the flares are detected are located in the saturated regime. Flares occur more frequently on the ZAMS stars having small $N_{\rm R}$. } \label{fig:NR_flare} 
\end{figure} 

\Figref{fig:NR_flare} shows the relationship between the flares and the rotation-activity. Most of the ZAMS stars in IC 2391 and IC 2602 on which the flares are detected are located in the saturated regime. Flares occur more frequently on the ZAMS stars having small $N_{\rm R}$. \citet{d16} show that GKM stars with small $N_{\rm R}$ show luminous flares. \citet{m20} examined the relationship between $N_{\rm R}$ and the flare rate for mid- to late-M dwarfs. They found a high flare rate for small $N_{\rm R}$ stars. Our results indicate that small $N_{\rm R}$ objects experience superflares, even for ZAMS stars. 

\section{Conclusion}
\label{conclusion}

We analyzed the {\it TESS} light curves for the $33$ ZAMS stars belonging to IC 2391 and IC 2602. Light variation was detected in all the ZAMS stars. This was considered to be caused by starspots. The amplitudes of the light curves range from $0.001-0.145\,\mathrm{mag}$. The starspot coverages range from $0.1-21\%$. 

\begin{enumerate}
\item The $R^{\prime}_{\lambda 8542}$ and of light curve amplitudes of ZAMS stars in IC 2391, IC 2602, and the Pleiades cluster are approximately two orders of magnitude larger than those of the Sun. These ZAMS stars are located on the extensions of the superflare stars and the Sun. This result suggests that superflare stars link the properties of the Sun to those of the ZAMS stars of ages between $30$ and $120\,\mathrm{Myr}$. 

\item The ZAMS stars in IC 2391, IC 2602, and the Pleiades cluster with small $N_{\rm R}$ show a light curve with a single frequency or possible shape change. The light curves have large amplitude, indicating inhomogeneous longitudinal distribution of large starspots. They also show saturated chromospheric Ca\, \emissiontype{II} emission lines, and have not reached the spin-down period. Most of the ﬂare events were detected on such objects in IC 2391 and IC 2602. Those enegies are more than $10^{33}\,\mathrm{erg}$, so that these flare events correspond to superﬂares.  

\item The ZAMS stars in IC 2391, IC 2602, and the Pleiades cluster with large $N_{\rm R}$ show beat or an complex light curve with a small amplitude. They have weak Ca\, \emissiontype{II} emission lines. The superflare events are rare on such objects in IC 2391 and IC 2602. 
\end{enumerate}


\begin{ack}
We wish to thank Dr. Notsu Y. and Dr. Namekata K. for comments. This paper includes data collected with the TESS mission, obtained from the MAST data archive at the Space Telescope Science Institute (STScI). M. Yamashita was supported by a scholarship from the Japan Association of University Women and would like to acknowledge them. This research was supported by a grant from the Hayakawa Satio Fund
awarded by the Astronomical Society of Japan. Y. I. is supported by JSPS KAKENHI grant number 17K05390.
\end{ack}

  

\end{document}